\documentclass[aps,prd,nofootinbib,onecolumn,a4paper]{revtex4-2}
\usepackage[colorlinks,linkcolor=blue,anchorcolor=violet,citecolor=red]{hyperref}
\usepackage{amsmath,amssymb}
\usepackage[subrefformat=parens]{subcaption}
\usepackage[dvipdfmx]{graphicx}
\usepackage{float}
\usepackage{comment}
\usepackage{xcolor}
\usepackage{bm,latexsym,amsmath,amssymb,amsfonts,mathrsfs}
\usepackage{bbold}
\usepackage{ascmac}
\usepackage{physics}
\usepackage{color,float}
\usepackage{enumerate}
\usepackage{braket}

\begin{document}
\newcommand{\simgt}{\lower.5ex\hbox{$\; \buildrel > \over \sim \;$}}
\newcommand{\simlt}{\lower.5ex\hbox{$\; \buildrel < \over \sim \;$}}
\def\barpsi{{\bar\psi}}
\title{\bf 
Violation of the two-time Leggett-Garg inequalities for a coarse-grained quantum field} 


\author{Masaki Tani}
\affiliation{Department of Physics,  Kyushu University, 744 Motooka, Nishi-Ku, Fukuoka 819-0395, Japan}

\author{Kosei Hatakeyama}
\affiliation{Department of Physics,  Kyushu University, 744 Motooka, Nishi-Ku, Fukuoka 819-0395, Japan}

\author{Daisuke Miki}
\affiliation{Department of Physics,  Kyushu University, 744 Motooka, Nishi-Ku, Fukuoka 819-0395, Japan}

\author{Yuki Yamasaki}
\affiliation{Department of Physics,  Kyushu University, 744 Motooka, Nishi-Ku, Fukuoka 819-0395, Japan}

\author{Kazuhiro Yamamoto}
\affiliation{Department of Physics,  Kyushu University, 744 Motooka, Nishi-Ku, Fukuoka 819-0395, Japan}
\affiliation{
Research Center for Advanced Particle Physics, Kyushu University, 744 Motooka, Nishi-ku, Fukuoka 819-0395, Japan}
\affiliation{
International Center for Quantum-field Measurement
Systems for Studies of the Universe and Particles (QUP),
KEK, Oho 1-1, Tsukuba, Ibaraki 305-0801, Japan}


\date{\today}

\begin{abstract}
We investigate the violation of the Leggett-Garg inequalities for a quantum field, focusing on the two-time quasi-probability distribution function of the dichotomic variable with a coarse-grained scalar field.
The Leggett-Garg inequalities are violated depending on the quantum state of the field and the size of coarse-graining. We also demonstrate that the violation of the Leggett-Garg inequalities appears even for the vacuum state and the squeezed state by properly constructing the dichotomic variable and the projection operator. 
\end{abstract}
\maketitle

\section{Introduction}
The Leggett-Garg inequalities were proposed to test the macrorealism to characterize classical systems \cite{Leggett85,Leggett02}, in which a macroscopic system is in a definite state at any given time in different available states and the state can be measured without any effect on the system. However, this can be violated in quantum systems as a result of the superposition principle and the state collapse. 
The Leggett-Garg inequalities utilize temporal correlations \cite{Emary14}, which are formulated in a similar analogy to the Clauser-Horne-Shimony-Holt (CHSH) inequality \cite{CHSC}, to test the spatial nonlocal correlation and the violation of the realism. 
The violation of the Leggett-Garg inequalities has been experimentally verified in many macroscopic quantum systems, e.g., in spin operators in qubit systems, superconducting circuits, and neutron interferometer  \cite{Ruskov,Palacios,Xu,Dressel,Knee,neutron}. 
The Leggett-Garg inequalities have also been applied to a test of the neutrino oscillations coherence \cite{neutrino}.
The quantum nature of gravitational interaction might be probed using the violation of the Leggett-Garg inequalities in the future \cite{MNY}.

Theoretical research on the Leggett-Garg inequalities is progressing (e.g., \cite{Moreira,Majidy19,Majidy21}).
In the present paper, we develop a theoretical formula for testing the violation of the Leggett-Garg inequalities in a quantum field theory. 
We utilize the two-time quasi-probability distribution function introduced in Ref.~\cite{Goldstein} and explored in Refs.~\cite{Halliwell2016,Halliwell2017,Halliwell2019,Halliwell2021}.
The present work is a generalization of the theoretical work for a harmonic oscillator in Ref.~\cite{Hatakeyama}, in which the violation of the two-time Leggett-Garg inequalities was investigated for various quantum states and projection operators. (For the studies on the violation of the Leggett-Garg inequalities in a harmonic oscillator, see also Refs.~\cite{Bose,Halliwell2022,Halliwell2023,Bose23}, cf.~Ref.~\cite{Asadian})
In Ref.~\cite{Hatakeyama}, a new technique to compute the two-time quasi-probability distribution function was developed. 
By generalizing the formulation to a quantum field, 
we demonstrate that the Leggett-Garg inequalities are violated for the dichotomic variable with a spatially coarse-grained quantum field, which will be useful to verify the quantum nature of a field.

The present paper is organized as follows: In Sec.~II, we briefly review the two-time quasi-probability distribution function and present the formulation for a quantum field theory  in $(3+1)$ dimensional Minkowski spacetime. We demonstrate that the violation of the Leggett-Garg inequalities appears for one mode coherent state in the quantum field, and the conditions for the violation are clarified. We discuss the effect of the squeezing of the quantum state of the field on the violation of the Leggett-Garg inequalities. 
We also discuss a non-trivial extension of the dichotomic variable and the projection operator, which reveals the violation of the Leggett-Garg inequalities for the vacuum state and the squeezed state. 
In Sec.~III, we demonstrate a similar violation for a 
 chiral massless field in  $(1+1)$ dimensional Minkowski spacetime.
Sec.~IV is devoted to summary and conclusions. In Appendix A, a brief summary of performing the integration of Eq.~(\ref{qssttsq}) is presented.

\section{LEGGETT-GARG INEQUALITIES}
We start with a brief review of the Leggett-Garg inequalities 
with the two-time quasi-probability distribution function. 
We introduce a dichotomic variable $Q$, which takes $\pm1$, and we assume that $Q_1$ and $Q_2$ are the values of $Q$ by measurement at the time $t_1$ and $t_2$, respectively. $s_1$ and $s_2$ are the numbers, $\pm1$, which we choose for the measurements at $t_1$ and $t_2$, respectively. 
Then, we have $(1+s_1Q_1)(1+s_2Q_2)\geq0$. 
Within the framework of macrorealism, there exists a joint probability function $p(Q_1,Q_2)$ to give the expectation values, 
which take the values of $0\leq p(Q_1,Q_2)\leq 1$, then the expectation value of $(1+s_1Q_1)(1+s_2Q_2)$
must be non-negative: 
\begin{align}
    \label{lgi}
    \langle(1+s_1Q_1)(1+s_2Q_2)\rangle\ge0.
\end{align}
This is a simple explanation of the two-time Leggett-Garg inequalities. Thus, depending on the choice of $s_1$ and $s_2$, we have four inequalities for the two times 
$t_1$ and $t_2$. 

In the quantum theory, the corresponding two-time Leggett-Garg inequalities are expressed regarding $\hat Q(t)$ as a Heisenberg operator of a dichotomic quantum variable, which gives $\pm1$ by a measurement. 
Corresponding variables $\hat Q_1$ and $\hat Q_2$ are defined by the results of measurements of 
${\hat Q}(t_1)=e^{i\hat Ht_1}\hat Qe^{-i\hat Ht_1}$ and
${\hat Q}(t_2)=e^{i\hat Ht_2}\hat Qe^{-i\hat Ht_2}$, respectively,
where we assume that the system evolves unitarily through the Hamiltonian $\hat H$. 
The two-time quasi-probability function is introduced by 
\begin{eqnarray}
&&q_{s_1,s_2}(t_1,t_2)=\frac{1}{8}{\rm Tr}[(1+s_1\hat Q(t_1))(1+s_2\hat Q(t_2))\rho_0]+(1\leftrightarrow2),
\end{eqnarray}
where $\rho_0$ is the initial density operator.
Introducing the projection operator, 
$\hat P_s=(1+s\hat Q)/2$, 
and its Heisenberg operator by
\begin{eqnarray}
&&{\hat P}_s(t)=e^{i\hat H t}{\hat P}_se^{-i\hat H t}={1\over 2}e^{i\hat H t}(1+s\hat Q)e^{-i \hat H t}
={1\over 2}(1+s\hat Q(t)), 
\end{eqnarray}
the quasi-probability distribution function is written as
\begin{eqnarray}
  &&q_{s_1,s_2}(t_1,t_2)={1\over 2}{\rm Tr}[\hat P_{s_1}(t_1)\hat P_{s_2}(t_2)\rho_0]+(1\leftrightarrow2)={\rm Re}
  {\rm Tr}[\hat P_{s_2}(t_2)\hat P_{s_1}(t_1)\rho_0].
\end{eqnarray}
We note that $q_{s_1,s_2}(t_1,t_2)$ satisfies the relations of the probability\cite{Halliwell2019}
\begin{eqnarray}
&&\langle \hat Q(t_1)\rangle =\sum_{s_1,s_2=\pm1} s_1q_{s_1,s_2}(t_1,t_2),\\
&&\langle \hat Q(t_2)\rangle =\sum_{s_1,s_2=\pm1} s_2q_{s_1,s_2}(t_1,t_2),\\
&&{1\over 2}\langle \{\hat Q(t_1),\hat Q(t_2)\}\rangle=\sum_{s_1,s_2=\pm1} s_1s_2q_{s_1,s_2}(t_1,t_2),
\end{eqnarray}
where $\{\hat Q(t_1),\hat Q(t_2)\}=\hat Q(t_1)\hat Q(t_2)+\hat Q(t_2)\hat Q(t_1)$. However, $q_{s_1,s_2}(t_1,t_2)$ may have negative values in quantum theory, then we call $q_{s_1,s_2}(t_1,t_2)$ quasi-probability.

The above formula can be applied to a continuous quantum variable of harmonic oscillator  \cite{Halliwell2021,Halliwell2022,Halliwell2023,Hatakeyama}.
In Ref. \cite{Hatakeyama} a useful formula to compute the quasi-probability distribution function is developed, which we apply to a quantum field in $(3+1)$ dimensional Minkowski spacetime. 
In the present paper, we consider a massless scalar field 
$\phi({x})$, expressed as 
\begin{align}
    \hat\phi(t, \bm x)&=
    \frac{1}{(2\pi)^{{3}/{2}}}
    \int d^{3}k\left(\frac{1}{\sqrt{2\omega_{ k}}}e^{-i\omega_k t+i\bm{k}\cdot\bm{x}}\hat a_{\bm{k}}+\frac{1}{\sqrt{2\omega_{k}}}e^{i\omega_k t-i\bm{k}\cdot\bm{x}}\hat a_{\bm{k}}^{\dagger}\right),
\end{align}
where $\hat a_{\bm k}$ and $\hat a_{\bm k}^\dagger$ are the annihilation and creation operators satisfying $[\hat a_{\bm k},\hat a_{\bm k'}^\dagger]=\delta^{(3)}(\bm k-\bm k')$, and $\omega_k=|\bm k|$. 
As the dichotomic operator, we adopt 
\begin{eqnarray}
    Q(t)=\rm{sgn}(\hat{\bar\phi}(t)-\varphi(t)),
\end{eqnarray}
and the operators $\hat{\bar\phi}(t)$ is defined 
by the coarse-grained field of $\hat\phi(\bm x,t)$ using the Gaussian window function with the scale $L$ as
\begin{align}
  \hat{\bar{\phi}}(t)=\frac{1}{\pi^{3/2}L^3}\int d^3x \hat\phi(\bm{x},t) e^{-{\bm{x}^2}/{L^2}},
\end{align}
where $\varphi(t)$ can be chosen
arbitrarily. We note that $\hat\phi(\bm x,t)$ should be understood as a Heisenberg operator.
Then, we may write
$
    \hat{\bar{\phi}}(t)=
     {(2\pi)^{-{3}/{2}}}
    \int d^{3}k(u_{\bm{k}}(t)\hat a_{\bm{k}}+u_{\bm{k}}^{*}(t)\hat a_{\bm{k}}^{\dagger})
$
with $u_{\bm{k}}(t)=
e^{-i\omega_k t-{\bm{k}^{2}L^{2}}/{4}}/\sqrt{2\omega_k}$.


The projection operator is given by
\begin{eqnarray}
  P_s(t)=\frac{1}{2}(1+s\times{\rm sgn}(\hat{\bar\phi}(t)-\varphi(t)))=\theta(s(\hat{\bar\phi}(t)-\varphi(t)),
\end{eqnarray}
where $\theta(z)$ is the Heaviside function.
The quasi-probability distribution function is
\begin{eqnarray}
  q_{s_1,s_2}(t_1,t_2)
  &=& {\rm Re~Tr}[ \theta(s_2(\hat {\bar\phi}(t_2)-\varphi(t_2)))\theta(s_1(\hat {\bar\phi}(t_1)-\varphi(t_2)))\rho_0].
\end{eqnarray}
The use of the mathematical formula $\theta'(z-c)=\delta(z-c)={(2\pi)}^{-1}\int_{-\infty}^{\infty}dpe^{-ip(z-c)}$ allows to write
\begin{eqnarray}
\theta(s(\hat {\bar\phi}(t)-\varphi(t)))=\int_0^\infty dc \int_{-\infty}^\infty {dp\over 2\pi} e^{ip(-s(\hat{\bar\phi}(t)-\varphi(t))+c)}.
\label{sxb}
\end{eqnarray}
When the initial state $\rho_0$ at $t=0$ is a pure state $\rho_0=|\psi_0\rangle\langle \psi_0|$,
we have
\begin{eqnarray}
  q_{s_1,s_2}(t_1,t_2)
  &=&{\rm Re}\left[
  \int_0^\infty \int_0^\infty dc_1 dc_2 \int_{-\infty}^\infty\int_{-\infty}^\infty\frac{dp_1dp_2}{(2\pi)^2}
  \langle \psi_0|   e^{-ip_2s_2\hat {\bar\phi}(t_2)}e^{-ip_1s_1\hat {\bar\phi}(t_1)}|\psi_0\rangle 
  e^{ip_2(c_2+s_2\varphi(t_2))} e^{ip_1(c_1+s_1\varphi(t_1))}\right].
  \nonumber\\
\end{eqnarray}
\subsection{Coherent state}
In this section, we consider the initial state
that a mode $\bm{\ell}$ is the coherent state and 
the other modes are the ground state, i.e., $\ket{\psi_0}=D_{\bm{\ell}}(\xi)\ket{0}$, where $\ket{0}$ denotes the vacuum state.  
We here define the displacement operator as
\begin{align}
    D_{\bm{\ell}}(\xi_{\bm{\ell}})=\exp(\xi_{\bm{\ell}}\hat a_{\bm{\ell}}^{\dagger}-\xi_{\bm{\ell}}^{*}\hat a_{\bm{\ell}}),
\end{align}
and we have
\begin{align}
    &D^{\dagger}_{\bm{\ell}}(\xi_{\bm{\ell}})e^{-ip_2s_2\hat {\bar\phi}(t_2)}e^{-ip_1s_1\hat {\bar\phi}(t_1)}D_{\bm{\ell}}(\xi_{\bm{\ell}})
=\exp(-\xi_{\bm{\ell}}\hat a_{\bm{\ell}}^{\dagger}+\xi_{\bm{\ell}}^{*}\hat a_{\bm{\ell}})\exp(-ip_2s_2 \int \frac{d^3k}{(2\pi)^{3/2}} (u_k(t_2)\hat a_{\bm{k}}+u_k^{*}(t_2)\hat a_{\bm{k}}^{\dagger})) \nonumber\\
&\hspace{5.5cm}\times \exp(-ip_1s_1 \int \frac{d^3k}{(2\pi)^{3/2}} (u_k(t_1)\hat a_{\bm{k}}+u_k^{*}(t_1)\hat a_{\bm{k}}^{\dagger}))\exp(\xi_{\bm{\ell}}\hat a_{\bm{\ell}}^{\dagger}-\xi_{\bm{\ell}}^{*}\hat a_{\bm{\ell}}).
\label{csf}
\end{align}
Using the BCH fomula, $e^{X+Y}=e^{X}e^{Y}e^{-[X,Y]/2},e^{X}e^{Y}=e^{[X,Y]}e^{Y}e^{X}$, which hold for the operators $X$ and $Y$ satisfying $[X,Y]={\rm constant}$, 
the quasi-probability function leads to
\begin{align}
    q_{s_1,s_2}(t_1,t_2)&={\rm Re}\biggl[\int_{0}^\infty\int_{0}^\infty{dc_1dc_2}\int_{-\infty}^\infty\int_{-\infty}^\infty \frac{dp_1dp_2}{(2\pi)^2}
    \exp\Bigl(-\frac{A}{2}(p_1^2+p_2^2)-p_1p_2s_1s_2B
        \notag\\    &\qquad
        {-}ip_1(s_1(E(t_1){-}\varphi(t_1))-c_1)
    -ip_2(s_2(E(t_2){-}\varphi(t_2))-c_2)\Bigr)\biggr],
    \label{equivalence}
\end{align}
where $A$ and $B$ are defined by
\begin{align}
    A&=\frac{1}{(2\pi)^{{3}}}\int d^3k\frac{1}{2\omega_k}e^{-{L^2 \bm{k}^2}/{2}} =\frac{1}{4L^2\pi^2}, \\
    B&=\frac{1}{(2\pi)^{{3}}}\int d^3k\frac{e^{-i\omega_k(t_2-t_1)}}{2\omega_k}e^{-{L^2 \bm{k}^2}/{2}}
    =\frac{1}{4L^2\pi^2}{-}i\sqrt{\frac{\pi}{2}}\frac{e^{-\frac{(t_2-t_1)^2}{2L^2}}}{4L^3\pi^2}(t_2-t_1)\left(1{{-}} {\rm erf}\left[\frac{i(t_2-t_1)}{\sqrt{2}L}\right]\right),
\end{align}
where ${\rm erf}(z)$ is the error function, 
and $E(t)$ is defined by 
\begin{align}
    \label{Et}
    E(t)&={1\over (2\pi)^{3/2}}
    \sqrt{\frac{2}{\omega_\ell}}|\xi_\ell|e^{-{L^2\bm{\ell}^2}/{4}}\cos(\omega_\ell t-\alpha_{\bm{\ell}})
\end{align}
with $\xi_{\bm \ell}=|\xi_{\bm{\ell}}|e^{i\alpha_{\bm \ell}}$. In the present paper, we assume $\alpha_{\bm \ell}=0$. 
From Eq.~(\ref{equivalence}), we note that the choice of the coherent state as the initial condition with $\varphi(t)=0$
is equivalent to the choice of the vacuum state as the initial condition by choosing $\varphi(t)=-E(t)$. Therefore, we assume $\varphi(t)=0$ in this section.

\begin{figure}[tbp]
    \centering
    \includegraphics[width=8.5cm]{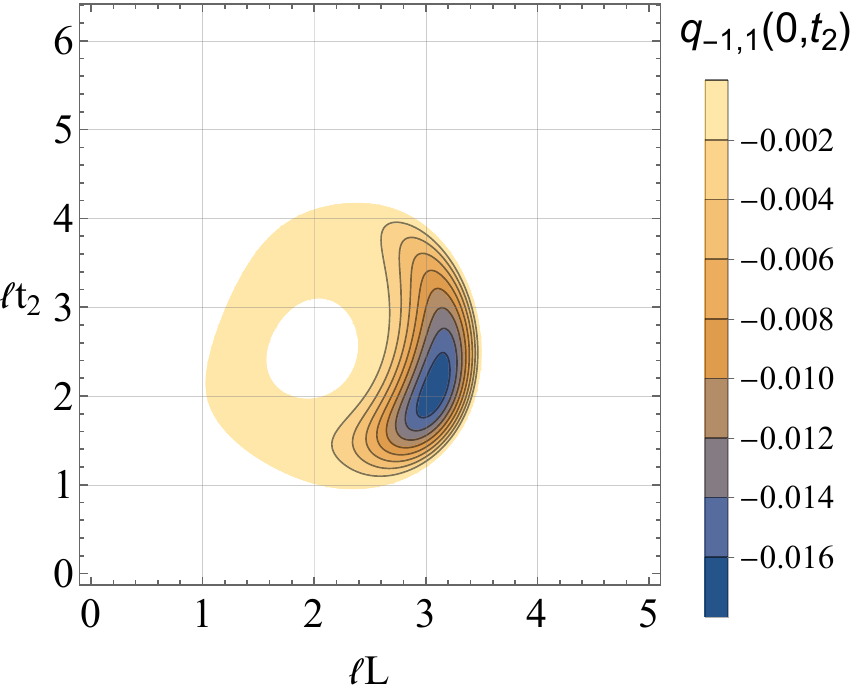}
    \hspace{0.4cm}
    \includegraphics[width=8.5cm]{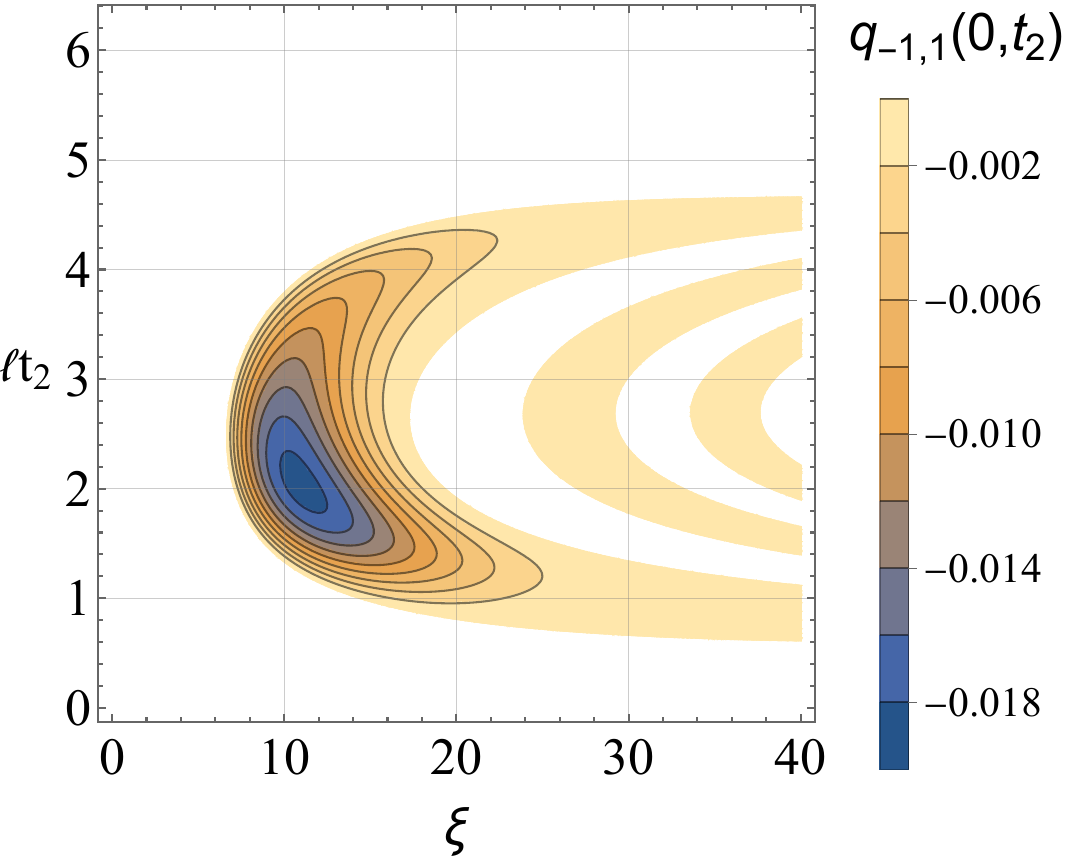}
    \caption{Left panel plots the contour of 
    $q_{-1,1}(0,t_2)$ on the plane of $\ell L$ and $\ell t_2$, where we fixed $\xi=8$.
    Right panel shows the contour $q_{-1,1}(0,t_2)$ on the plane of $\xi$ and $\ell t_2$, where we fixed $\ell L=10/3$. In these panels, the quasi-probability takes negative values in the colored regions, where the Leggett-Garg inequality is violated. 
    }
    \label{fig:encoher}
\end{figure}
By performing the Gaussian integral in Eq.~(\ref{equivalence}), we have
\begin{eqnarray}
     &&q_{s_1,s_2}(t_1,t_2)={\rm Re}\biggl[\frac{1}{2\pi\sqrt{A^2-B^2}}\int_0^{\infty}\int_0^{\infty}dc_1dc_2\exp\biggr(-\frac{1}{2(A^2-B^2)}\Bigl\{A(c_1^2+c_2^2)-2Bs_1s_2c_1c_2
          \nonumber\\&&\qquad\qquad
-2s_1(AE(t_1)-BE(t_2))c_1 
-2s_2(AE(t_2)-BE(t_1))c_2+A(E^2(t_1)+E^2(t_2))-2BE(t_1)E(t_2)\Bigr\}\biggl)\biggr],
\label{qsstta}
\end{eqnarray}
%
which can be performed numerically, as demonstrated in Figure~\ref{fig:encoher}.
The appearance of the Leggett-Garg inequalities depends on the parameter of the amplitude of the coherent state $\xi(=\xi_{\bm \ell})$ and on the size of the coarse-graining $L$ of the field through $E(t)$, $A$, and $B$. 
Fig.~\ref{fig:encoher} plots the contour of the quasi-probability distribution function $q_{-1,1}(0,t_2)$.  
The left panel of Fig.~\ref{fig:encoher} 
plots $q_{-1,1}(0,t_2)$ on the plane of $\ell L$ and $\ell t_2$ with fixed $\xi=8$, where $\ell=|\bm \ell|$. 
The colored regions show the ones where the quasi-probability distribution function has negative values. Similarly, the right panel of Fig.~\ref{fig:encoher} shows $q_{-1,1}(0,t_2)$ on the plane of $\xi$ and $\ell t_2$, where $\ell L$ is fixed as $\ell L=10/3$. 
These panels demonstrate that the two-time Leggett-Garg inequality with $s_1=-1$ and $s_2=1$ is violated 
when we choose the parameters $\xi$ and $L$ properly.
The results show that the optimized values are $\ell L\simeq\pi$ and $\xi\simeq 10$ for the violation. 
Thus the clear violation of the Leggett-Garg inequality occurs for the choice of $s_1=-1$ and $s_2=1$. 
We found a violation for the case $s_1=-1$ and $s_2=-1$, but the violation is very weak compared with the case $s_1=-1$ and $s_2=1$. 

The behaviors of Fig.~\ref{fig:encoher}
can be understood as follows. 
We note that 
$L$ is the parameter to represent the size of the coarse-graining of the field. From the left panel of Fig.~\ref{fig:encoher}, $L$ must be taken to be $1<\ell L<3.5$
for the violation of the Leggett-Garg inequality. If $L$ is much larger than the wavelength of the mode in the coherent state, $L\ell\gg1$, the coherence in $\hat {\bar \phi}(t)$  is washed out by the coarse-graining of the field. 
Conversely, if $L$ is much shorter than the wavelength of the coherent state, $L\ell\ll1$, the vacuum fluctuations of shorter-wavelength modes in $\hat \phi(t)$ dominate a measured value of $\hat {\bar \phi}(t)$ over the quantum nature of the mode 
${\bm \ell}$ in the coherent state.  
The optimized value is $\ell L\simeq \pi$, which means that $L$ equals the half of wavelength of the coherent wave $\pi/\ell$.  
From the right panel of Fig.~\ref{fig:encoher}, 
we see that $\xi$, the magnitude of the coherent state parameter must be greater than $7$ for the violation.
The optimized value is $\xi\simeq 10$, and then the violation of the Leggett-Garg inequalities gets weaker as $\xi$ becomes larger. 

The quasi-probability distribution function for a harmonic oscillator in a coherent state 
is periodic with respect to $t_1$ and $t_2$ with the period of the harmonic oscillator \cite{Hatakeyama}. This is because the evolution of the system is unitary. In contrast, the quasi-probability distribution function for the coarse-grained quantum field
shows no periodic behavior. 
The evolution of our system is unitary too, but this feature would be understood as an analogy of the decoherence effect coming from the fact that the dichotomic variable for the field theory is constructed by the coarse-graining of the field. 
This would be understood as a result of the quantum correlation in the field.

\subsection{Squeezed state with a coherent mode}
In this section, we consider the initial state  as the two-mode squeezed coherent state defined by
\begin{align}
    \ket{\psi_0}=
    D_{\bm{\ell}}(\xi_{\bm{\ell}})\Bigl[\prod
    _{\substack{{\bm k}\\ {\rm (independent)}}}
    S_{\bm{k},-\bm{k}}(\zeta_{\bm{k}})\ket{0}\Bigr],
\end{align}
where $S_{\bm{k},-\bm{k}}(\zeta_{\bm{k}})$ is the two-mode squeezed operator defined as
$
    S_{\bm{k},-\bm{k}}(\zeta_{\bm{k}})
    =\exp(\zeta^{*}_{\bm{k}}a_{\bm{k}}a_{-\bm{k}}-\zeta_{\bm{k}}a^{\dagger}_{\bm{k}}a^{\dagger}_{-\bm{k}}).
$  
We use the properties of the 
the squeezing operator transforms annihilation and creation operators as follows,
\begin{align}
    &{S}_{\bm{k},-\bm{k}}^{\dagger}(\zeta_{\bm{k}})\hat{a}_{\bm{k}} {S}_{\bm{k},-\bm{k}}(\zeta_{\bm{k}})
    =\hat{a}_{\bm{k}}\cosh r_{\bm{k}}-\hat{a}^{\dagger}_{-\bm{k}}e^{i\theta_{\bm{k}}}\sinh r_{\bm{k}},\\
    &{S}_{\bm{k},-\bm{k}}^{\dagger}(\zeta_{\bm{k}})\hat{a}_{\bm{k}}^{\dagger} {S}_{\bm{k},-\bm{k}}(\zeta_{\bm{k}})
    =\hat{a}_{\bm{k}}^{\dagger}\cosh r_{\bm{k}}-\hat{a}_{-\bm{k}}e^{-i\theta_{\bm{k}}}\sinh r_{\bm{k}},
\end{align}
where we used $\zeta_{\bm k}=r_{\bm k}e^{i\theta_{\bm k}}$.
After a similar computation to the previous section, we have the following expression for the quasi-probability
\begin{eqnarray}
    q_{s_1,s_2}(t_1,t_2)&=&
    {\rm Re}\int_{0}^{\infty}\int_{0}^{\infty}dc_1dc_2
    \int_{-\infty}^{\infty}\int_{-\infty}^{\infty}\frac{dp_1dp_2}{(2\pi)^2}\exp\biggl[-\frac{A_{\text{sq}}(t_1)p^2_1+A_{\text{sq}}(t_2)p^2_2}{2}-p_1p_2s_1s_2B_{\text{sq}}(t_1,t_2)
    \nonumber\\
    &&~~~~~~~~-ip_1(s_1(E(t_1){-}\varphi(t_1))-c_1)-ip_2(s_2(E(t_2){-}\varphi(t_2))-c_2)\biggr],
    \label{qps}
\end{eqnarray}
where we defined 
\begin{align}
    &A_{\text{sq}}(t)=
    \frac{1}{(2\pi)^{\color{blue}{3}}}\int d^3k\frac{e^{-{L^2k^2}/{2}}}{2\omega_k}
    \left(\cosh{2r_{\bm{k}}}-\sinh{2r_{\bm{k}}}\cos{(2\omega_kt-\theta_{\bm{k}})}\right),
    \label{defAA}\\
    &B_{\text{sq}}(t_1,t_2)=\frac{1}{(2\pi)^{\color{blue}{3}}}\int d^3 k
    \frac{e^{-{L^2k^2}/{2}}}{2\omega_k}
    \left(e^{-i\omega_k(t_2-t_1)}\cosh^2{r_{\bm{k}}}-\sinh{2r_{\bm{k}}}\cos{(\omega_k(t_1+t_2)-\theta_{\bm{k}})}{{+}}e^{i\omega_k(t_2-t_1)}\sinh^2{r_{\bm{k}}}
    \right),
    \label{defBB}
\end{align}
and $E(t)$ is defined by Eq.~(\ref{Et}). 
Here we note that assuming the squeezed state with a coherent mode 
as the initial condition with $\varphi(t)=0$
is equivalent to assuming the squeezed state as the initial condition with $\varphi(t)=-E(t)$. We assume $\varphi(t)=0$ hereafter in this section. The case of the coherence state is reproduced by choosing $r_{\bm k}=0$.
\begin{figure}[tbp]
    \centering
    \centering
    \includegraphics[width=8.5cm]{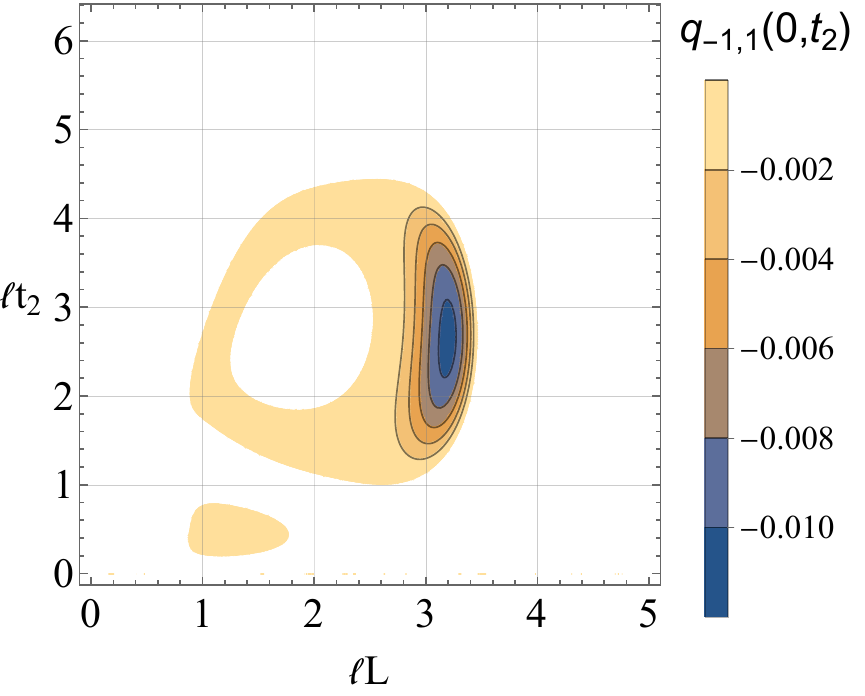}
    \hspace{0.4cm}
    \includegraphics[width=8.5cm]{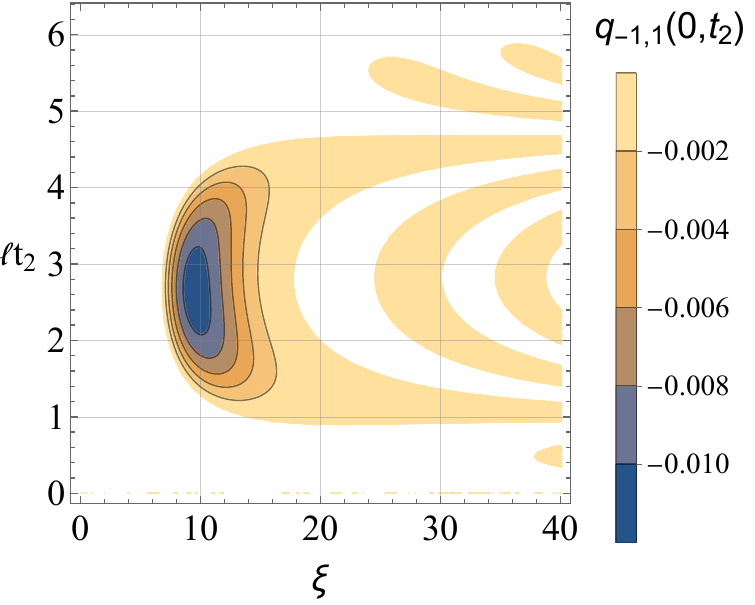}
    \caption{
    Contour of $q_{-1,1}(0,t_2)$ 
    for the squeezed state with a coherent mode.
    The left panel shows the contour $q_{-1,1}(0,t_2)$ on the plane of $\ell L$ and $\ell t_2$, where we fixed $\xi=8$ and $r=0.5$. 
    The right panel shows the contour plots on the plane of $\xi$ and $\ell t_2$, 
    where we fixed $\ell L=10/3$ and $r=0.5$.  
    }
    \label{fig:ensq}
\end{figure}

When $r_{\bm{k}}$ does not depend on the mode $\bm k$ and $\theta_{\bm{k}}=0$, i.e., $r_{\bm k}=r$, $A_{\text{sq}}(t)$ and 
$B_{\text{sq}}(t_1,t_2)$ reduce to
\begin{align}
    &A_{\text{sq}}(t)=
\frac{Le^{-2r}-it{\sqrt{2\pi}e^{-2t^2/L^2}}{\rm erf}\Bigl({i\sqrt{2}t}/{L}\Bigr)\sinh{2r}}{4\pi^2 L^3} \\
    &B_{\text{sq}}(t_1,t_2)=
\frac{1}{{4\sqrt{2}\pi^{2} L^3}} \biggr\{ \sqrt{2}e^{-2r}L-i\sqrt{\pi}(t_2-t_1)e^{-(t_2-t_1)^2/{2L^2}}\left(1-{\rm erf}\left(\frac{i(t_2-t_1)}{\sqrt{2}L}\right)\cosh{2r}\right)
\nonumber\\
&\hspace{3.5cm}
+i\sqrt{\pi}(t_1+t_2)e^{-(t_1+t_2)^2/{2L^2}}{\rm erf}\left(\frac{i(t_1+t_2)}{\sqrt{2}L}\right)\sinh{2r}\biggr\}.
\end{align}
After integration of the right-hand side of Eq.(\ref{qps}) over $p_1$ and $p_2$, we have
\begin{eqnarray}
    q_{s_1,s_2}(t_1,t_2)&=&
    \text{Re} \biggl[   \frac{1}{2\pi\sqrt{A_{\text{sq}}(t_1)A_{\text{sq}}(t_2)-B_{\text{sq}}(t_1,t_2)^2}}
    \int_0^{\infty}\int_0^{\infty}dc_1dc_2
    \exp\biggl(-\frac{1}{2(A_{\text{sq}}(t_1)A_{\text{sq}}(t_2)-B_{\text{sq}}(t_1,t_2)^2)}
    \nonumber\\
   &&\biggl\{A_{\text{sq}}(t_2)c_1^2+A_{\text{sq}}(t_1)c_2^2-2s_1s_2B_{\text{sq}}(t_1,t_2)c_1c_2 -2c_1s_1\Bigl(A_{\text{sq}}(t_2)E(t_1)-B(t_1,t_2)E(t_2)\Bigr)
   \nonumber\\
   &&
   -2c_2s_2\Bigl(A_{\text{sq}}(t_1)E(t_2)-B_{\text{sq}}(t_1,t_2)E(t_1)\Bigr)+A_{\text{sq}}(t_2)E(t_1)^2+A_{\text{sq}}(t_1)E(t_2)^2\nonumber \\
   &&-2B_{\text{sq}}(t_1,t_2)E(t_1)E(t_2)\biggr\}\biggr)\biggr].
   \label{qssttsq}
\end{eqnarray}
Figure~\ref{fig:ensq} shows the same as Fig.~\ref{fig:encoher} but for the squeezed state with a coherent mode 
Eq.~(\ref{qssttsq}) with $r=0.5$.
The left panel plots the contour of $q_{-1,1}(0,t_2)$ on the plane of $\ell L$ and $\ell t_2$ with fixed $\xi=8$, while the right panel shows the contour on the plane of $\xi$ and $\ell t_2$ with fixed $\ell L=10/3$. 
The overall behavior of $q_{-1,1}(0,t_2)$ in Fig.~\ref{fig:ensq} is similar to that of Fig.~\ref{fig:encoher}, though the effect of the squeezed initial state changes the pattern of the contour. The effect of squeezing increases the region where the Leggett-Garg inequality is violated.
Especially, the periodic behavior appears in time $t_2$ when $\xi$ is large, but it is not very clear because the violation is weak there.

\begin{figure}[t]
    \centering
    \includegraphics[width=10cm]{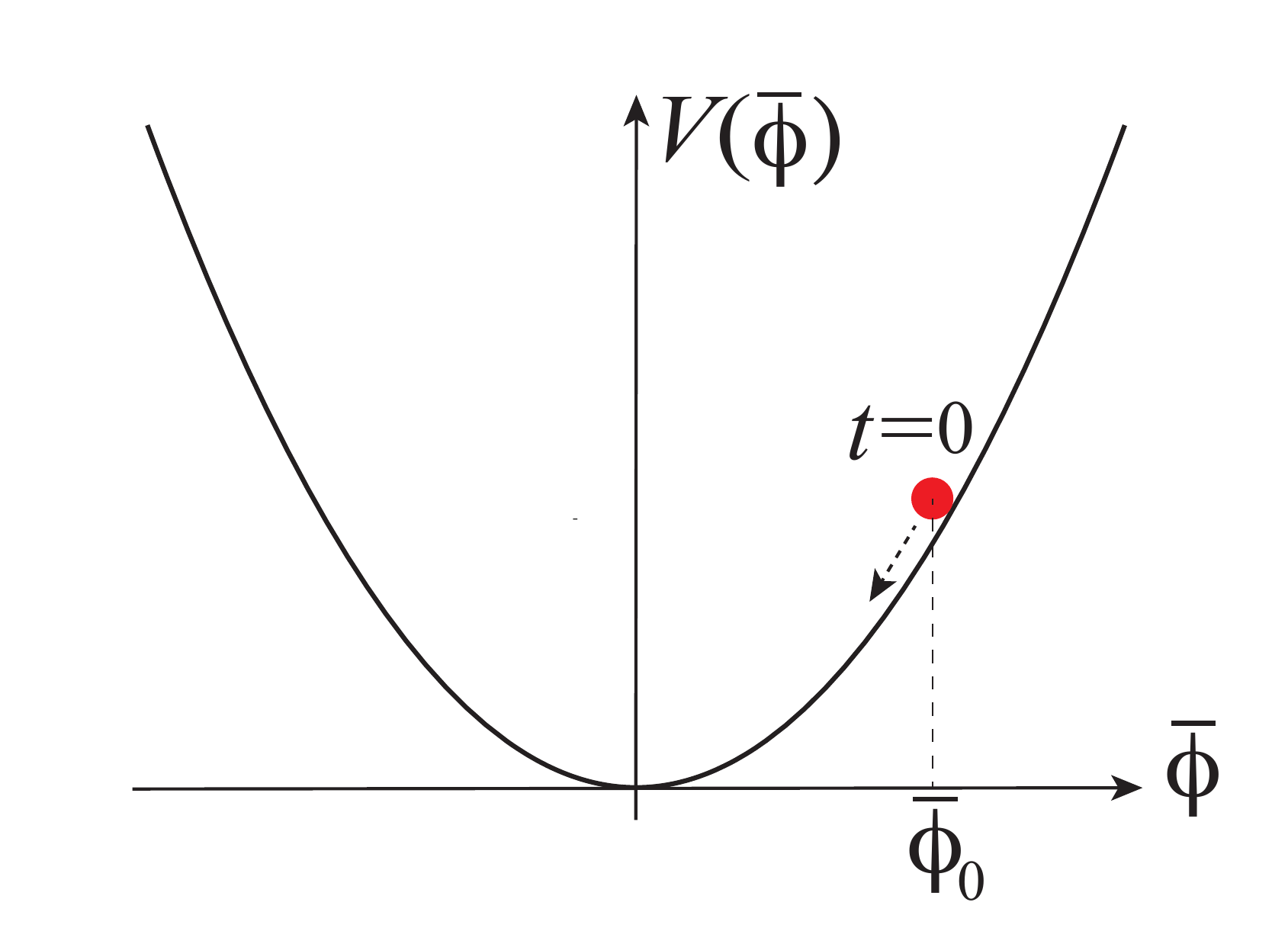}
    \caption{
    A schematic plot of the potential $V(\bar \phi)$ for $\bar\phi$. The violation of the Leggett-Garg inequality can be understood using an analogy of the dynamics of $\bar\phi$ as a harmonic oscillator.  }
    \label{fig:potential}
\end{figure}
\subsection{Intuitive understanding of the violation}

Let us discuss the reason why the violation of  the Leggett-Garg inequalities appears. We investigated the Leggett-Garg inequalities in a quantum field theory, but it is useful to consider it in an analogy with a harmonic oscillator \cite{Hatakeyama}. 
The expectation value of the variable $\hat {\bar\phi}(t)$ is given by
\begin{eqnarray}
{\rm Tr}[\hat {\bar\phi}(t)\rho_0]=
{1\over (2\pi)^{3/2}}    \sqrt{\frac{2}{\omega_\ell}}|\xi_\ell|e^{-{L^2{\bm \ell}^2}/{4}}\cos(\omega_\ell t-\alpha_{{\bm \ell}})=E(t).
\label{expEd}
\end{eqnarray}
As we assumed  $\alpha_{\bm \ell}=0$ in the present paper, then the expectation value is ${\rm Tr}[\hat {\bar\phi}(t)\rho_0]=\bar \phi_0 \cos \omega_\ell t$, where $\bar\phi_0=|\xi_{\bm \ell}|e^{-L^2{\bm \ell}^2/4}/\sqrt{4\pi^{3}\omega_\ell}>0$. Figure \ref{fig:potential} shows a schematic plot of the potential for $\bar\phi$ in analogy with a harmonic oscillator, which gives the solution $\bar \phi_0 \cos \omega_\ell t$. 
As we also assumed $t_1=0$, then the red mark in the figure denotes the initial position of $\bar\phi$ at $t_1=0$, then $\bar\phi$ starts to move to the left. As shown in the previous section, the violation of the Leggett-Garg inequalities clearly occurs for $s_1=-1$ and $s_2=1$. This means that the violation appears for the quasi-probability that the measurement at $t_1$ gives $\bar\phi<0$ and the measurement $t_2$ gives $\bar\phi>0$. 
The results of these measurements are opposite to the expected values of 
${\rm Tr}[\hat {\bar\phi}(t)\rho_0]=\bar\phi_0\cos(\omega_\ell t)$. 
Namely the expectation values at $t_1$ and $t_2$
are ${\rm Tr}[\hat {\bar\phi}(t_1=0)\rho_0]>0$ and ${\rm Tr}[\hat {\bar\phi}(t_2)\rho_0]<0$ for $\pi/2<\omega_\ell t_2<3\pi/2$. 
Because ${\rm Tr}[\hat {\bar\phi}(t)\rho_0]=\bar \phi_0\cos(\omega_\ell t)$ can be regarded as a classically expected motion, therefore the violation of the Leggett-Garg inequalities occurs when the measurements give the opposite values to the classically expected values. This occurs in quantum mechanics because the wave function has a spread due to the superposition principle, which is the origin of the violation of realism. 

\subsection{Projection operator for the vacuum state and the squeezed state}
\begin{figure}[t]
        \centering 
    \includegraphics[width=8.5cm]{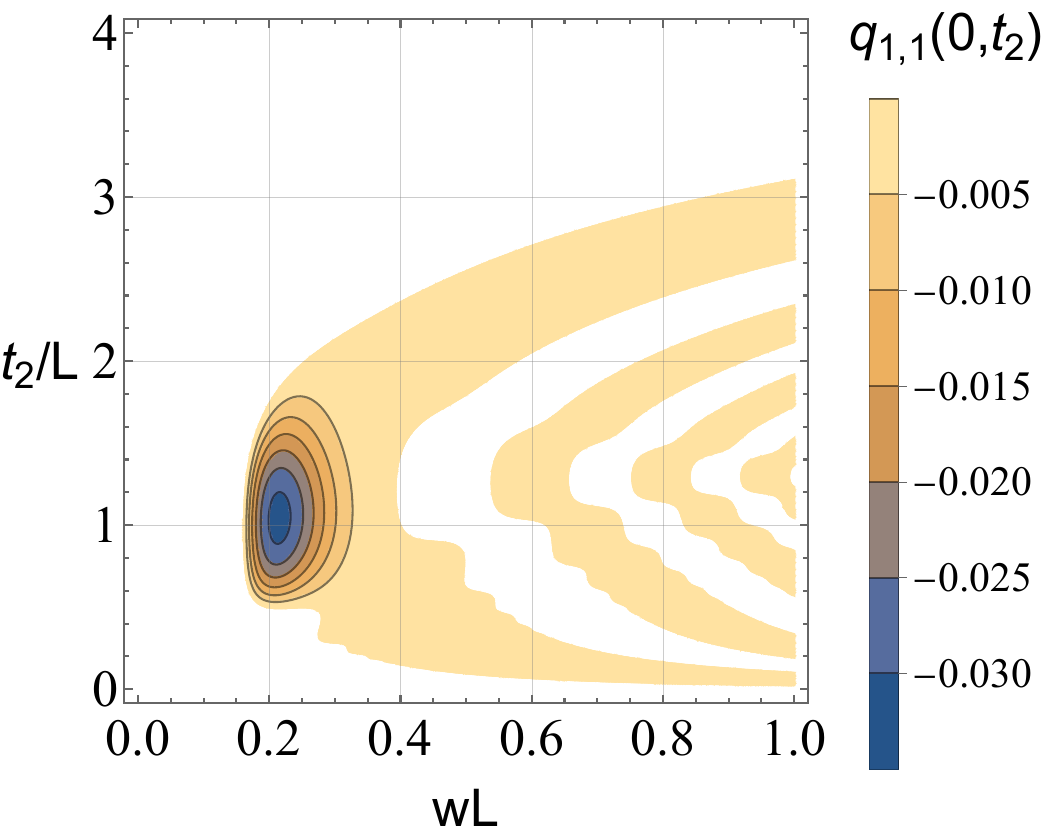}
   \hspace{0.4cm}
    \includegraphics[width=8.5cm]{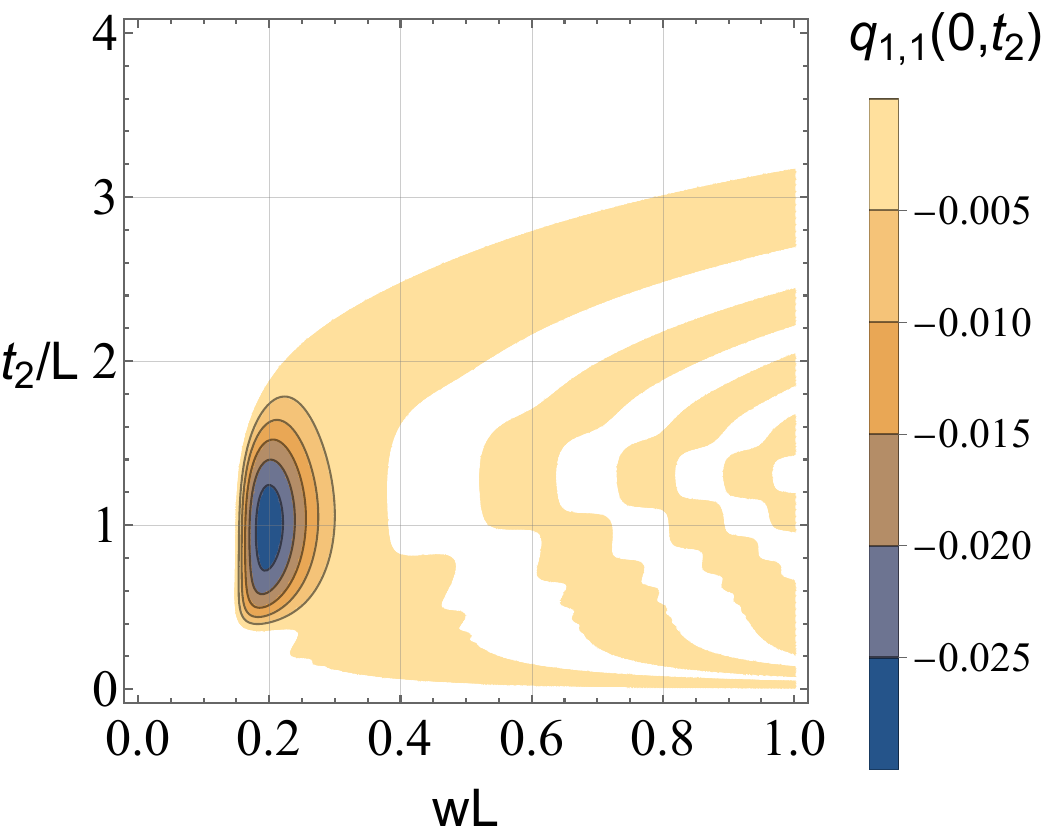}
        \caption{Contour plot of $q_{1,1}(0,t_2)$ for the squeezed state with Eq.~(\ref{Pstthree}). 
        The left (right) panel shows the contour $q_{1,1}(0,t_2)$ on the plane of $wL$ and $t_2/L$, where we fixed
       $r=0.3$ ($r=0.5$) and $\theta_0=0$. 
       In these panels, the quasi-probability has negative values in the colored regions, where the Leggett-Garg inequality is violated.
        \label{fig:box}}
\end{figure}
\begin{figure}[t]
        \centering 
         \includegraphics[width=8.5cm]{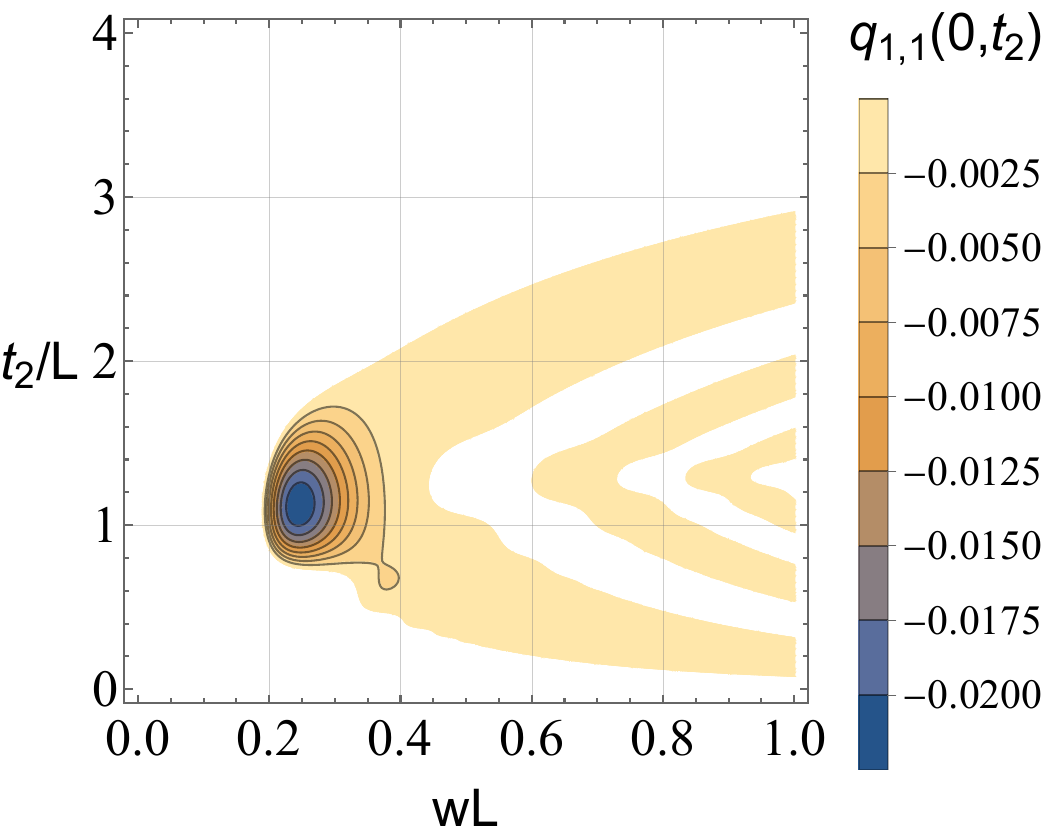}
        \caption{
        Same as the panels in Fig. \ref{fig:box} but for the case $r=0$  and $\theta_0=0$, the vacuum state. 
        \label{fig:boxv}}
\end{figure}
In this subsection, we adopt 
\begin{align}
    Q(t)=1+\textrm{sgn}(\hat{\bar\phi}(t)-w)+\textrm{sgn}(-\hat{\bar\phi}(t)-w)
\end{align}
as a dichotomic operator,
where the operator $\hat{\bar\phi}(t)$ is defined
by the coarse-grained field  using the Gaussian window function with the scale $L$ in a similar way to the previous subsection.
Here $w$ is a parameter. 
The projection operator is given by
\begin{align}
    P_s(t)&=\frac{1}{2}(1+s+\textrm{sgn}(\hat{\bar\phi}(t)-w)+\textrm{sgn}(-\hat{\bar\phi}(t)-w)=\theta(s(\hat{\bar\phi}(t)-w))+\theta(-s(\hat{\bar\phi}(t)+w))+\frac{1}{2}(s-1),
    \label{Pstthree}
\end{align}
where $\theta(z)$ is the Heaviside function.
This dichotomic variable defined by the above projection operator is
understood as follows. When the absolute value of a result of a measurement of the coarse-grained field $\hat {\bar \phi}(t)$ is larger than $w$, we assign $Q=1$. On the other hand,
the absolute value of a result of a measurement of $\hat {\bar \phi}(t)$ is smaller than $w$, we assign $Q=-1$. 
Therefore, the projection operator $P_s$ with $s=1$ gives the projection onto the region $|\bar \phi(t)|>w$, while $P_s$ with $s=-1$
does the projection onto the region $|\bar \phi(t)|\leq w$.

The quasi-probability function is given by evaluating
\begin{align}
    q_{s_1,s_2}(t_1,t_2)
    &= {\rm Re~Tr}\bigg[\left\{ \theta(s_2(\hat{\bar\phi}(t_2)-w))+\theta(-s_2(\hat{\bar\phi}(t_2)+w))+\frac{1}{2}(s_2-1)\right\}\nonumber \\
    &\hspace{33pt}\times\left\{\theta(s_1(\hat{\bar\phi}(t_1)-w))+\theta(-s_1(\hat{\bar\phi}(t_1)+w))+\frac{1}{2}(s_1-1)\right\} \rho_0\bigg].
\end{align}
Using the mathematical formula $\theta'(z-a)=\delta(z-a)=\frac{1}{2\pi}\int_{-\infty}^{\infty}dpe^{-ip(z-a)}$,
we have
\begin{eqnarray}
    q_{s_1,s_2}(t_1,t_2)
    &=&{\rm Re}\biggl[\langle \psi_0|\left\{\int_0^\infty dc_2\int_{-\infty}^\infty \frac{dp_2}{2\pi}\left(e^{-ip_2(s_2(\hat{\bar\phi}(t_2)-w-c_2)}+e^{ip_2(s_2(\hat{\bar\phi}(t_2)+w)+c_2)}\right)+\frac{1}{2}(s_2-1)\right\}\nonumber \\
    &&\hspace{27pt}\times\left\{\int_0^\infty dc_1\int_{-\infty}^\infty \frac{dp_1}{2\pi}\left(e^{-ip_1(s_1(\hat{\bar\phi}(t_1)-w)-c_1)}+e^{ip_1(s_1(\hat{\bar\phi}(t_1)+w)+c_1)}\right)+\frac{1}{2}(s_1-1)\right\}|\psi_0\rangle\biggr],
\end{eqnarray}
where we assumed that the initial state $\rho_0$ at $t=0$ is the squeezed state $\rho_0=|\psi_0\rangle\langle \psi_0|$ with 
\begin{align}
    \ket{\psi_0}=\biggl[
    \prod
    _{\substack{{\bm k}\\ {\rm (independent)}}}
    S_{\bm{k},-\bm{k}}(\zeta_{\bm{k}})\biggr]\ket{0}.
\end{align}
After similar calculations to those in the previous subsection,
we have the quasi-probability distribution function
\begin{align}
    q_{s_1,s_2}(t_1,t_2)&={\rm Re}\biggl[2\int_{0}^{\infty}\int_{0}^{\infty}dc_1dc_2
    \int_{-\infty}^{\infty}\int_{-\infty}^{\infty}\frac{dp_1dp_2}{(2\pi)^2}\nonumber \\
    &\hspace{23pt}\exp\left\{-\frac{A_{\text{sq}}(t_1)p^2_1+A_{\text{sq}}(t_2)p^2_2}{2}-p_1p_2s_1s_2B_{\text{sq}}(t_1,t_2)+ip_1(s_1w+c_1)+ip_2(s_2w+c_2)\right\}\nonumber \\
    &+2\int_{0}^{\infty}\int_{0}^{\infty}dc_1dc_2
    \int_{-\infty}^{\infty}\int_{-\infty}^{\infty}\frac{dp_1dp_2}{(2\pi)^2}\nonumber \\
    &\hspace{23pt}\exp\left\{-\frac{A_{\text{sq}}(t_1)p^2_1+A_{\text{sq}}(t_2)p^2_2}{2}+p_1p_2s_1s_2B_{\text{sq}}(t_1,t_2)+ip_1(s_1w+c_1)+ip_2(s_2w+c_2)\right\}\nonumber \\
    &+2\int_0^\infty dc_1\int_{-\infty}^\infty \frac{dp_1}{2\pi}\exp{-\frac{A_{\text{sq}}(t_1)p^2_1}{2}+ip_1(s_1w+c_1)}\times\frac{1}{2}(s_2-1)\nonumber \\
    &+2\int_0^\infty dc_2\int_{-\infty}^\infty \frac{dp_2}{2\pi}\exp{-\frac{A_{\text{sq}}(t_2)p^2_2}{2}+ip_2(s_2w+c_2)}\times\frac{1}{2}(s_1-1)+\frac{1}{4}(s_1-1)(s_2-1)\biggr],
    \label{qpw}
\end{align}
where $A_{\rm sq}(t)$ and $B_{\rm sq}(t_1,t_2)$ are defined by Eqs. (\ref{defAA}) and (\ref{defBB}), respectively.
Here we assume that $r_{\bm{k}}$ does not depend on $\bm k$ and $\theta_{\bm{k}}=\theta_0=0$, i.e., $r_{\bm k}=r$.
Eq.~(\ref{qpw}) can be evaluated in a similar way to the previous subsection,
\begin{eqnarray}
    q_{s_1,s_2}(t_1,t_2)&=&
    \text{Re} \biggl[   \frac{1}{\pi\sqrt{A_{\text{sq}}(t_1)A_{\text{sq}}(t_2)-B_{\text{sq}}(t_1,t_2)^2}}
    \int_0^{\infty}\int_0^{\infty}dc_1dc_2
    \exp\biggl(-\frac{1}{2(A_{\text{sq}}(t_1)A_{\text{sq}}(t_2)-B_{\text{sq}}(t_1,t_2)^2)}
    \nonumber\\
   &&\hspace{10pt}\biggl\{A_{\text{sq}}(t_2)c_1^2+A_{\text{sq}}(t_1)c_2^2-2s_1s_2B_{\text{sq}}(t_1,t_2)c_1c_2 +2c_1s_1w\Bigl(A_{\text{sq}}(t_2)-B(t_1,t_2)\Bigr)
   \nonumber\\
   &&\hspace{10pt}
   +2c_2s_2w\Bigl(A_{\text{sq}}(t_1))-B_{\text{sq}}(t_1,t_2)\Bigr)+\Bigl(A_{\text{sq}}(t_2)+A_{\text{sq}}(t_1)-2B_{\text{sq}}(t_1,t_2)\Bigr)w^2\biggr\}\biggr)\nonumber\\
   &&\hspace{7pt}+\frac{1}{\pi\sqrt{A_{\text{sq}}(t_1)A_{\text{sq}}(t_2)-B_{\text{sq}}(t_1,t_2)^2}}
    \int_0^{\infty}\int_0^{\infty}dc_1dc_2
    \exp\biggl(-\frac{1}{2(A_{\text{sq}}(t_1)A_{\text{sq}}(t_2)-B_{\text{sq}}(t_1,t_2)^2)}
    \nonumber\\
   &&\hspace{10pt}\biggl\{A_{\text{sq}}(t_2)c_1^2+A_{\text{sq}}(t_1)c_2^2+2s_1s_2B_{\text{sq}}(t_1,t_2)c_1c_2 +2c_1s_1w\Bigl(A_{\text{sq}}(t_2)+B(t_1,t_2)\Bigr)
   \nonumber\\
   &&\hspace{10pt}
   +2c_2s_2w\Bigl(A_{\text{sq}}(t_1))+B_{\text{sq}}(t_1,t_2)\Bigr)+\Bigl(A_{\text{sq}}(t_2)+A_{\text{sq}}(t_1)+2B_{\text{sq}}(t_1,t_2)\Bigr)w^2\biggr\}\biggr)\nonumber\\
   &&\hspace{7pt}+\frac{1}{2}(s_2-1)\left(1-{\rm erf}\left[\frac{s_1w}{\sqrt{2A_{\text{sq}}(t_1)}}\right]\right)+\frac{1}{2}(s_1-1)\left(1-{\rm erf}\left[\frac{s_2w}{\sqrt{2A_{\text{sq}}(t_2)}}\right]\right)+\frac{1}{4}(s_1-1)(s_2-1)\biggr].\nonumber\\
\end{eqnarray}

Figure \ref{fig:box} plots the contour of $q_{1,1}(0,t_2)$ on the plane of $w L$ and $ t_2/L$, where the left (right) panel adopted $r=0.3$ ($r=0.5$). 
One can see that the quasi-probability distribution function has the negative values smaller than $-0.03$ ($-0.025$) for $r=0.3$ ($r=0.5$) at $Lw\simeq 0.2$.
Figure \ref{fig:boxv} plots the same as Fig.~\ref{fig:box} but for the case $r=0$, the vacuum state.
For the vacuum state and the squeezed states, the expectation values of $\hat {\bar \phi}(t)$ are always zero. On the other hand, 
$q_{1,1}(0,t_2)$ denotes the quasi-probability that the measurement at $t_1=0$ gives $|\bar \phi(t_1)|>w$ and 
 the measurement at $t_2$ gives $|\bar \phi(t_2)|>w$. 
 This is the counter-intuitive result of measurements against the expectation values. 
 This occurs because of a spread of the wave function, coming from the superposition principle of the quantum mechanical systems.

\section{One-dimensional chiral massless scalar field}

In this section, we consider a chiral massless quantum field in (1+1) dimensional Minkowski spacetime motivated by Ref.~\cite{Hotta}.
We consider a massless scalar field, which is expressed as 
\begin{align}
    \hat\phi(t, x)&=
    \frac{1}{(2\pi)^{{1}/{2}}}
    \int_0^\infty dk\left(\frac{1}{\sqrt{2{k}}}e^{-i\omega_k (t+x)}\hat a_{{k}}+\frac{1}{\sqrt{2{k}}}e^{i\omega_k (t+x)}\hat a_{{k}}^{\dagger}\right),
\end{align}
where $\hat a_{k}$ and $\hat a_{k}^\dagger$ are the annihilation and creation operators satisfying $[\hat a_{k},\hat a_{k'}^\dagger]=\delta^{}(k-k')$, and $\omega_k=k$. 
As the dichotomic operator, we adopt 
\begin{eqnarray}
    Q(t)=\rm{sgn}(\hat{\bar\phi}{}'(t)-\varphi(t)),
\end{eqnarray}
where we defined the operators $\hat{\bar\phi}{}'(t)$ 
by the coarse-grained quantity of $\hat\phi'(x,t)$ using the Gaussian window function with the scale $L$ as
\begin{align}
  \hat{\bar{\phi}}{}'(t)=\frac{1}{\sqrt{\pi}L}\int_{-\infty}^\infty dx \hat\phi'({x},t) {e^{-x^2/L^2}},
\end{align}
where $\phi'(t,x)=\partial\phi(x,t)/\partial x$, and $\varphi(t)$ can be chosen arbitrarily. 
Then, we may have
\begin{eqnarray}
    \hat{\bar{\phi}}{}'(t)=
     {1\over \sqrt{2\pi}}
    \int_0^\infty dk(u_{{k}}(t)\hat a_{{k}}+u_{{k}}^{*}(t)\hat a_{{k}}^{\dagger})
    \end{eqnarray}
with $u_{{k}}(t)=ik
e^{-ik t-{{k}^{2}L^{2}}/{4}}/\sqrt{2k}$.
The projection operator and the quasi-probability distribution  function are given by
\begin{eqnarray}
  &&P_s(t)=\frac{1}{2}(1+s\times{\rm sgn}(\hat{\bar\phi}{}'(t)-\varphi(t)))=\theta(s(\hat{\bar\phi}{}'(t)-\varphi(t))),
\\
  &&q_{s_1,s_2}(t_1,t_2)
  = {\rm Re~Tr}[ \theta(s_2(\hat {\bar\phi}{}'(t_2)-\varphi(t_2)))\theta(s_1(\hat {\bar\phi}{}'(t_1)-\varphi(t_2)))\rho_0].
\end{eqnarray}

\subsection{Coherent state}
\begin{figure}[tbp]
    \centering
    \includegraphics[width=8.3cm]{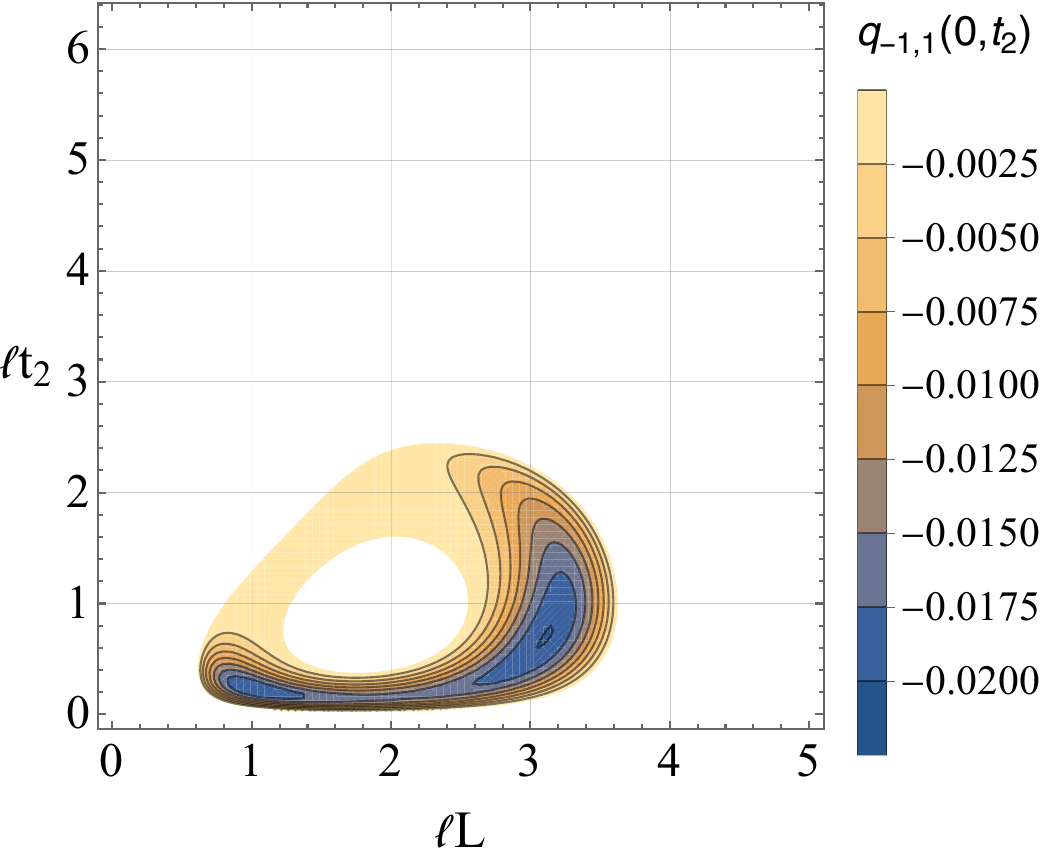}
    \hspace{0.7cm}
    \includegraphics[width=8.3cm]{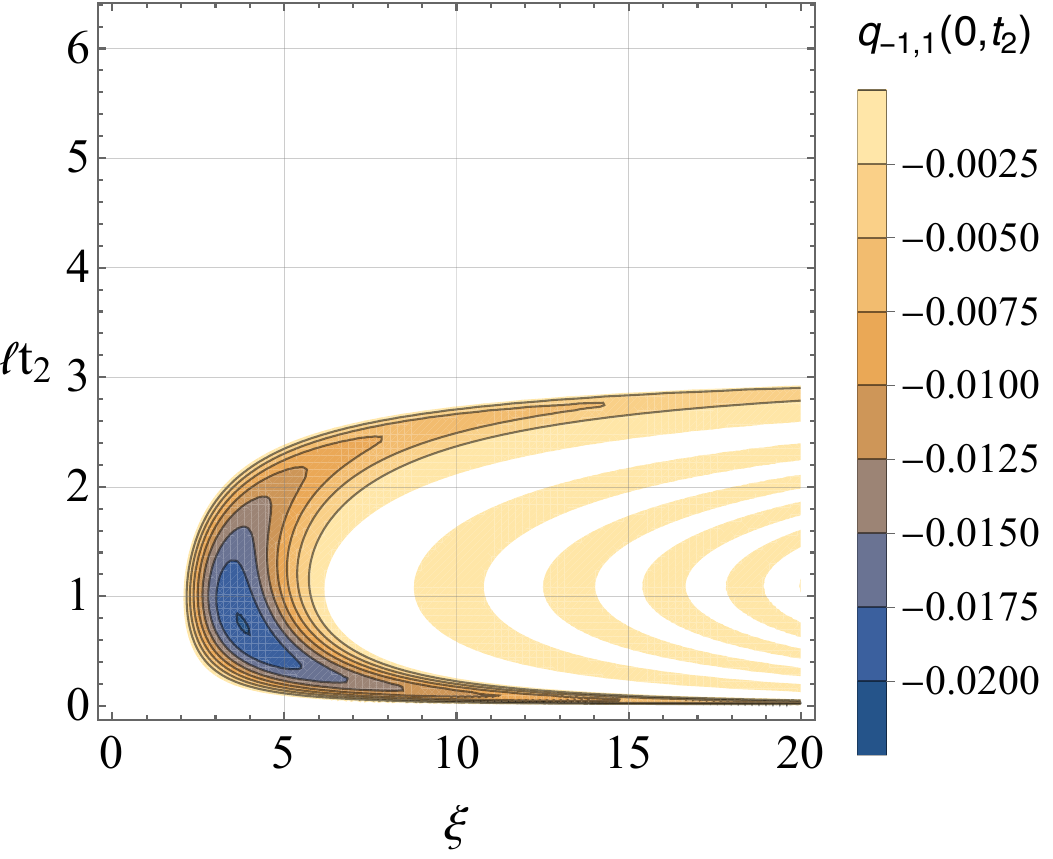}
    \caption{Contour of $q_{-1,1}(0,t_2)$ 
    for the one-mode coherent state of the field.
        The left panel plots the contour on the plane of $\ell L$ and $\ell t_2$, 
    where we fixed $\xi=3$. 
    The right panel plots the contour on the plane of $\xi$ and $\ell t_2$, 
    where we fixed $\ell L=10/3$. 
 In these panels, the quasi-probability takes negative values in the colored regions, where the Leggett-Garg inequality is violated. 
    }
    \label{fig:encoherone}
\vspace{0.5cm}
    \centering
    \includegraphics[width=8.5cm]{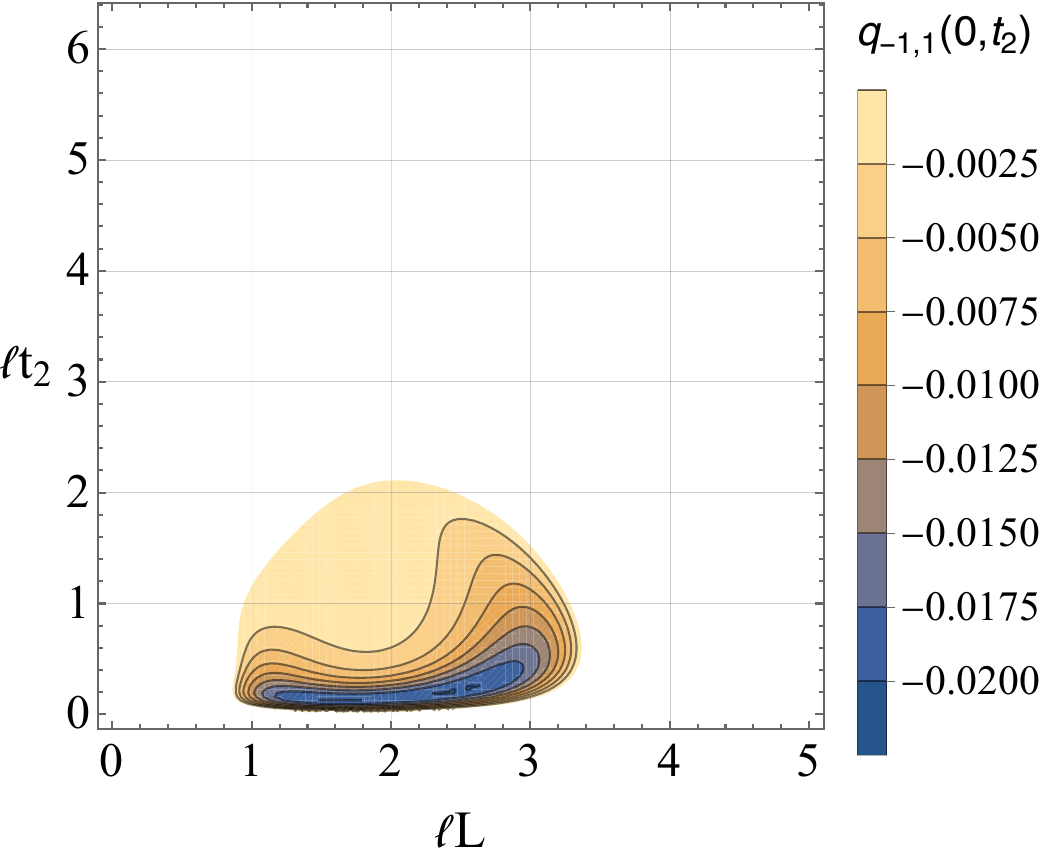}
    \hspace{0.4cm}
    \includegraphics[width=8.5cm]{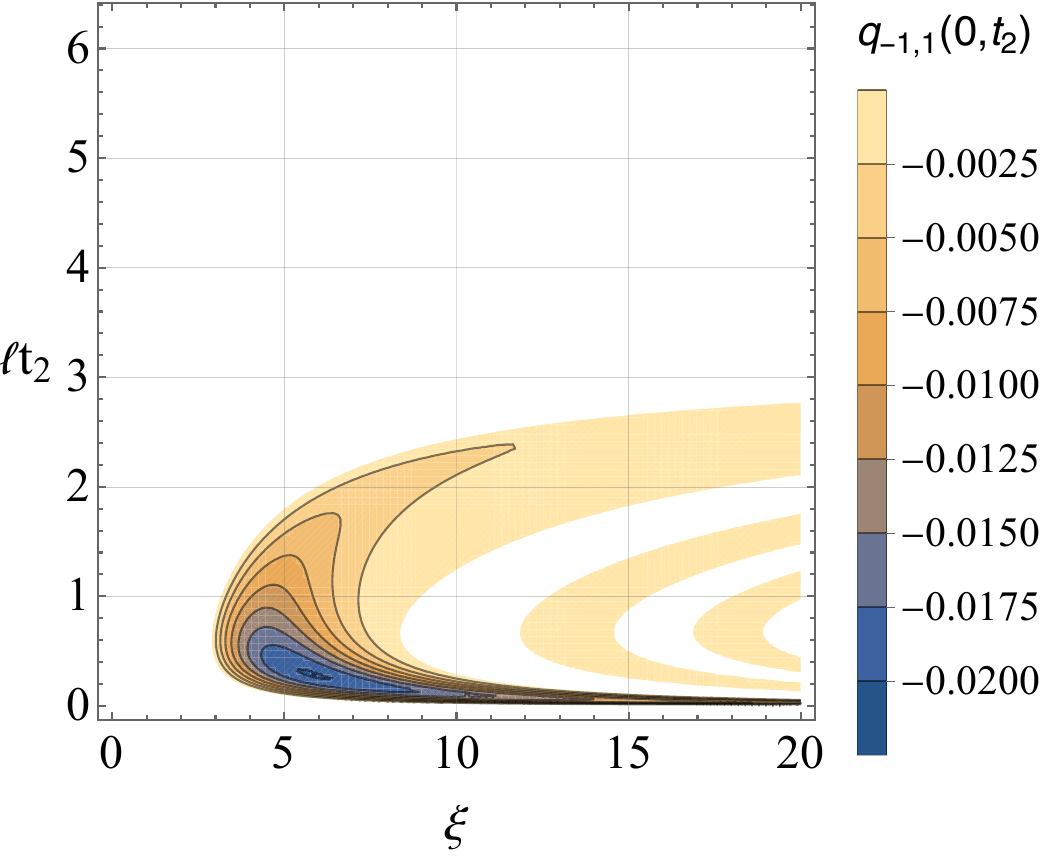}
    \caption{
    Same as Fig.~\ref{fig:encoherone} but
    for the squeezed field with a coherent mode 
    with $r_k=r=0.5$ and $\theta_k=0$.
    The left panel shows the contour $q_{-1,1}(0,t_2)$ on the plane of $\ell L$ and $\ell t_2$, where we fixed $\xi=3$.
    The right panel shows $q_{-1,1}(0,t_2)$ on the plane of $\xi$ and $\ell t_2$, 
    where we fixed $\ell L=10/3$.  
    }
    \label{fig:ensqone}
\end{figure}

We first consider the initial state
that a mode ${\ell}$ is in the coherent state and 
the other modes are the vacuum state, i.e., $\ket{\psi_0}=D_{{\ell}}(\xi)\ket{0}$. 
In this case, we have the same 
expression for the quasi-probability distribution function as Eq.~(\ref{qsstta}) 
but with replacing $A$, $B$, and $E(t)$  by
\begin{align}
 &A=\frac{1}{2\pi^{}}\int_0^\infty{e^{-{L^2 {k}^2}/{2}}\frac{k^2}{2k}} d k=\frac{1}{4 \pi L^2},
 \label{Aone}\\
&B=\frac{1}{2\pi^{}}\int_0^\infty{e^{-{L^2 {k}^2}/{2}}\frac{e^{-ik(t_1-t_2)}}{2k}} k^2d {k}=\frac{1}{4\pi L^2}{{-}}\frac{i}{4\pi L^3}e^{-\frac{(t_2-t_1)^2}{2L^2}}(t_2-t_1)\sqrt{\frac{\pi}{2}}\left[1{{-}}{\rm erf}\left( \frac{i(t_2-t_1)}{\sqrt{2}L}\right)\right],
\label{Bone}
\end{align}
and
\begin{align}
    &E(t)=
    \sqrt{\frac{\omega_\ell}{\pi}}|\xi_{\ell}|e^{-{L^2\ell^2}/{4}}\sin(\alpha_{\ell}-\omega_\ell t)
    \label{onedEt}
\end{align}
with $\xi=|\xi_{\ell}|e^{i\alpha_\ell}$.
We assume $\alpha_{\ell}=0$ in the present paper.
The quasi-probability distribution function can be evaluated numerically, as demonstrated in Figure~\ref{fig:encoherone}, which
plots the contour of the quasi-probability distribution function $q_{-1,1}(0,t_2)$ on the plane of $\ell L$ and $\ell t_2$ (left panel)
and on the plane of $\xi$ and $\ell t_2$ (right panel).
Similar behaviors as those in Fig.~\ref{fig:encoher} can be read. 
The appearance of the Leggett-Garg inequalities depends on the parameters of the amplitude of the coherent state $\xi$ and on the size of the coarse-graining $L$ of the field through $E(t)$, $A$, and $B$.  
The left panel of Fig.~\ref{fig:encoherone} 
plots the contour on the plane of $\ell L$ and $\ell t_2$ with fixed $\xi=3$. 
The colored regions show the one where the quasi-probability distribution function has negative values. 
Similarly, the right panel of Fig.~\ref{fig:encoherone} shows the contour on the plane of $\xi$ and $\ell t_2$, where $\ell L$ is fixed as $\ell L=10/3$.
The clear violation of the two-time Leggett-Garg inequalities appears for $s_1=-1$ and $s_2=1$ when we choose the parameters $\xi$ and $L$ properly. 
The results show that the optimized values are $\ell L\simeq\pi$ and $\xi\simeq4$. 

\subsection{Squeezed state with a coherent mode}
We next consider the squeezed state with a coherent mode
defined by
$
    \ket{\psi_0}=
    D_{\ell}(\xi_{\ell})[\prod_{k>0}S_k(\zeta_{k})]\ket{0},
    $
where $S_k(\zeta_{k})$ is the one-mode squeezing operator defined as
\begin{align}
S_k(\zeta_{k})
=\exp\left[
\frac{\zeta^{*}_{k}\hat a_{k}^2-\zeta_{k}\hat a^{\dagger}_{k}{}^2}{2}
\right].
\end{align}
In this case, we have the same expression for the quasi-probability distribution function as Eq.~(\ref{qps}) but with replacing $A_{\rm sq}(t)$ and $B_{\rm sq}(t_1,t_2)$ by 
\begin{align}
  &A_{\text{sq}}(t)=\frac{1}{2\pi} \int_0^{\infty} dk  e^{-{L^2k^2}/{2}}\frac{k^2}{2k}\bigg{(}\cosh{2r_k}-\sinh{2r_k}\cos{\big{(}2kt-\theta_k\big{)}}\bigg{)},
  \label{Asqone}\\
  &B_{\text{sq}}(t_1,t_2)=\frac{1}{2\pi} \int_0^{\infty} dk  e^{-{L^2k^2}/{2}}\frac{k^2}{2k}\bigg{(}\cosh{2r_k}\cos{\big{(}k (t_1-t_2) \big{)}} -\sinh{2r_k}\cos{\big{(}k( t_1+t_2) -\theta_k \big{)}}+i\sin(k(t_1-t_2)) \bigg{)},
  \label{Bsqone}
\end{align}
and $E(t)$ defined by Eq.~(\ref{onedEt}), where $\zeta_k=r_ke^{i\theta_k}$. 
Eqs. (\ref{Asqone}) and (\ref{Bsqone}) yield
\begin{align}
  &A_{\text{sq}}(t)=\frac{L(\cosh{2r}-\cos{\theta}\sinh{2r})+e^{-2t^2/L^2}\sqrt{2\pi}t(-i\cos{\theta}\text{erf}(i\sqrt{2}t/L)-\sin{\theta})\sinh{2r}}{4\pi L^3}
  \label{AsqoneB}\\
  &B_{\text{sq}}(t_1,t_2)=\frac{1}{4\sqrt{2}L^3\pi}\biggl\{-i\sqrt{\pi}(t_2-t_1)+\cosh{2r}\biggl(\sqrt{2}L-ie^{-(t_2-t_1)^2/2L^2}\sqrt{\pi}(t_2-t_1)\text{erf}\Bigl[-\frac{i(t_2-t_1)}{\sqrt{2}L}\Bigr]\biggr)
  \nonumber\\
  &\hspace{2cm}-\sinh{2r}\biggl(\sqrt{2}L\cos{\theta}+ie^{-(t_2+t_1)^2/2L^2}\sqrt{\pi}(t_2+t_1)\Bigl(\cos{\theta}\text{erf}\Bigl[\frac{i(t_2+t_1)}{\sqrt{2}L}\Bigr]-i\sin{\theta}\Bigr)\biggr)\biggr\}
  \label{BsqoneB}
\end{align}
when $\zeta_{k}$ does not depend on $k$, i.e., $\zeta_{k}=r$  and $\theta_k=\theta$. In this case,
we have the same expression for the quasi-probability distribution function as Eq.~(\ref{qssttsq}) but with $A_{\rm sq}(t)$, $B_{\rm sq}(t_1,t_2)$, and $E(t)$ defined by  
Eqs.~(\ref{AsqoneB}), (\ref{BsqoneB}), and (\ref{onedEt}), respectively.
%

Panels of Fig.~\ref{fig:ensqone} show the contour of the quasi-probability distribution function $q_{-1,1}(0,t_2)$. 
The left panel plots the contour of $q_{-1,1}(0,t_2)$ on the plane of $\ell L$ and $\ell t_2$
with fixed $\xi=3$, on the other hand, the right panel shows the contour on the plane of $\xi$ and $\ell t_2$ with fixed $\ell L=10/3$. 
In Fig.~\ref{fig:ensqone}, we assumed $r_k=0.5$ and $\theta_k=0$. 
Similarly to Fig.~\ref{fig:encoherone}, 
$q_{-1,1}(0,t_2)$ in Fig.~\ref{fig:ensqone} has negative values in the colored region in each panel. 
The overall behaviors of $q_{-1,1}(0,t_2)$ in Fig.~\ref{fig:ensqone} is similar to those of Fig.~\ref{fig:encoherone}, though the effect of the squeezed initial state changes the pattern of the contour. 

We can understand when the violation of the Leggett-Garg inequalities 
appears in a similar way to the three-dimensional case.
For the one-dimensional chiral field, in which the expectation value of the coarse-grained field is
\begin{eqnarray}
{\rm Tr}[\hat {\bar\phi}{}'(t)\rho_0]=
    \sqrt{\frac{\omega_\ell}{\pi}}|\xi_{\ell}|e^{-{L^2\ell^2}/{4}}\sin(\alpha_{\ell}-\omega_\ell t)
    =E(t).
    \label{onedEtd}
\end{eqnarray}
As we assumed $\alpha_\ell=0$, 
${\rm Tr}[\hat {\bar\phi}{}'(t)\rho_0]$ takes negative values for $0<\omega_\ell t<\pi$. On the other hand, $q_{-1,1}(0,t_2)$ is the quasi-probability that the measurement of $\hat {\bar\phi}{}'(t)$ at $t_2$ gives a positive value, which is opposite to the expectation values for $0<\omega_\ell t<\pi$, where 
the violation of the Leggett-Garg inequality appears as demonstrated in Figs. \ref{fig:encoherone} and \ref{fig:ensqone}.

\begin{figure}[tbp]
    \centering
    \includegraphics[width=8.5cm]{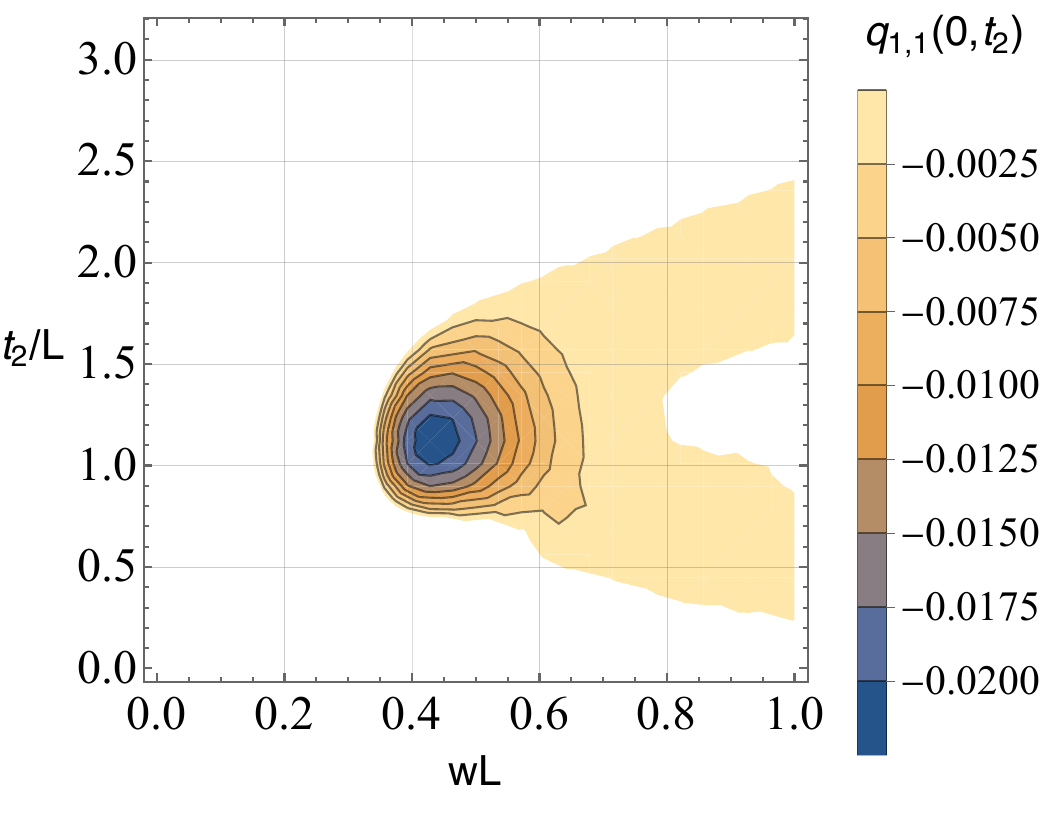}
    \hspace{0.5cm}
    \includegraphics[width=8.5cm]{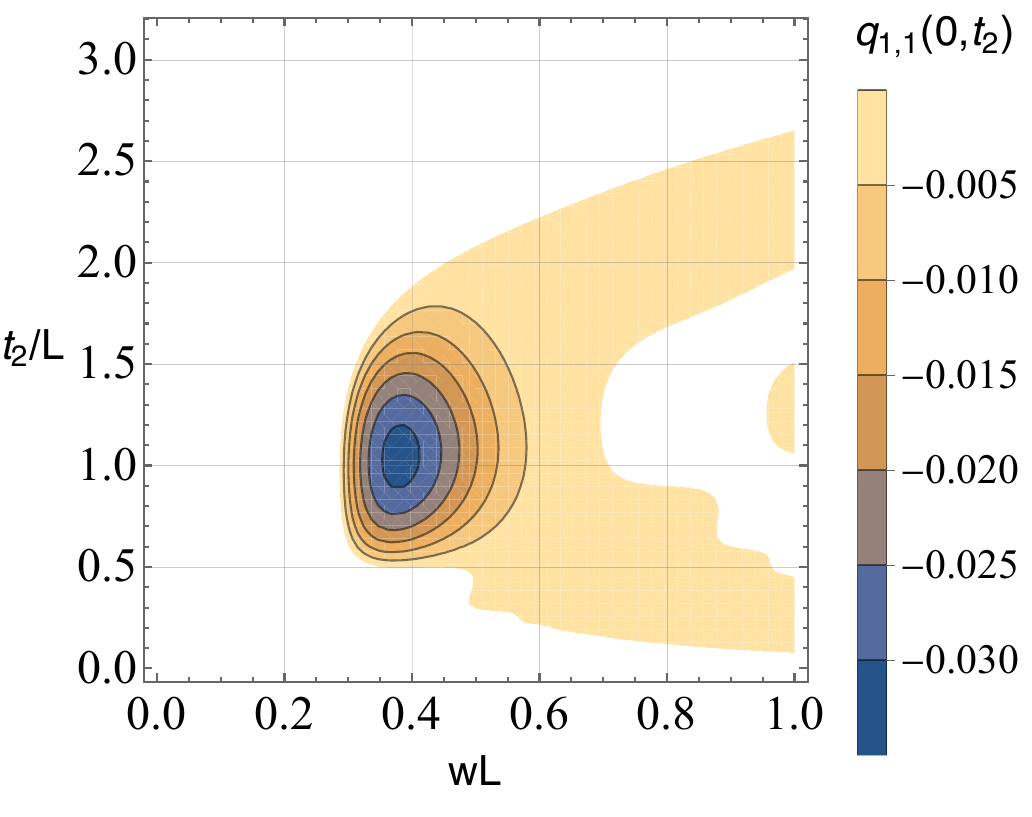}
    \caption{Contour of $q_{1,1}(0,t_2)$ 
    for the one-dimensional chiral massless field with
    the projection  operator Eq.~(\ref{PSoned}) on the plane of $wL$ and $t_2/L$, 
    where we fixed $r=0$ (left panel) and $r=0.3$ (right panel). 
 In these panels, the quasi-probability takes negative values in the colored regions, where the Leggett-Garg inequality is violated. 
}
    \label{fig:yamasakiond}
\end{figure}
\begin{figure}[tbp]
    \centering
    \includegraphics[width=8.5cm]{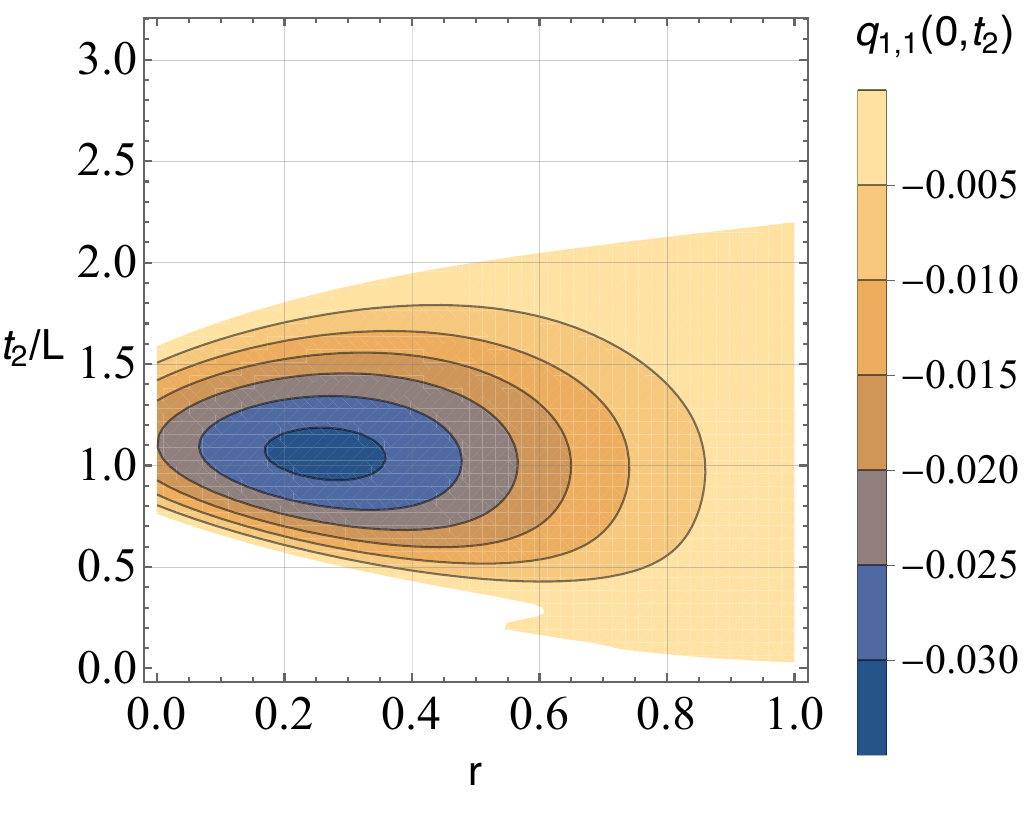}
    \caption{Contour of $q_{1,1}(0,t_2)$ 
    for the one-dimensional chiral massless field with
    the projection  operator Eq.~(\ref{PSoned}) 
      on the plane of squeezed parameter $r$ and $t_2/L$, 
    where we fixed $wL=0.4$.
}
    \label{fig:yamasakiondB}
\end{figure}

\subsection{Projection operator for the squeezed state and the vacuum state}
Similarly to Sec. II C, let us now consider the case when the dichotomic variable $Q(t)$ is adopted as  
\begin{align}
    Q(t)=1+\textrm{sgn}(\hat{\bar{\phi^\prime}}(t)-w)+\textrm{sgn}(-\hat{\bar{\phi^\prime}}(t)-w),
\end{align}
in which the projection operator is written as
\begin{align}
\label{PSoned}
    P_s(t)&=\frac{1}{2}(1+s+\textrm{sgn}(\hat{\bar\phi}^\prime(t)-w)+\textrm{sgn}(-\hat{\bar\phi}^\prime(t)-w)=\theta(s(\hat{\bar\phi}^\prime(t)-w))+\theta(-s(\hat{\bar\phi}^\prime(t)+w))+\frac{1}{2}(s-1).
\end{align}
The dichotomic variable is defined so that we assign
we assign $Q=1$ ($Q=-1$) when the absolute value of a result of a measurement of the coarse-grained field $\hat {\bar \phi}^\prime(t)$ is larger (smaller) than $w$. 
For the squeezed state defined by
$   \ket{\psi_0}=\prod_{k>0}S_k(\zeta_{k})\ket{0}$,
the expression of the quasi-probability distribution function is the same as 
Eq.~(\ref{qpw}) but with $A_{\rm sq}(t)$, $B_{\rm sq}(t_1,t_2)$, and $E(t)$ in this section.  

\begin{figure}[t]
    \centering
    \includegraphics[width=8cm]{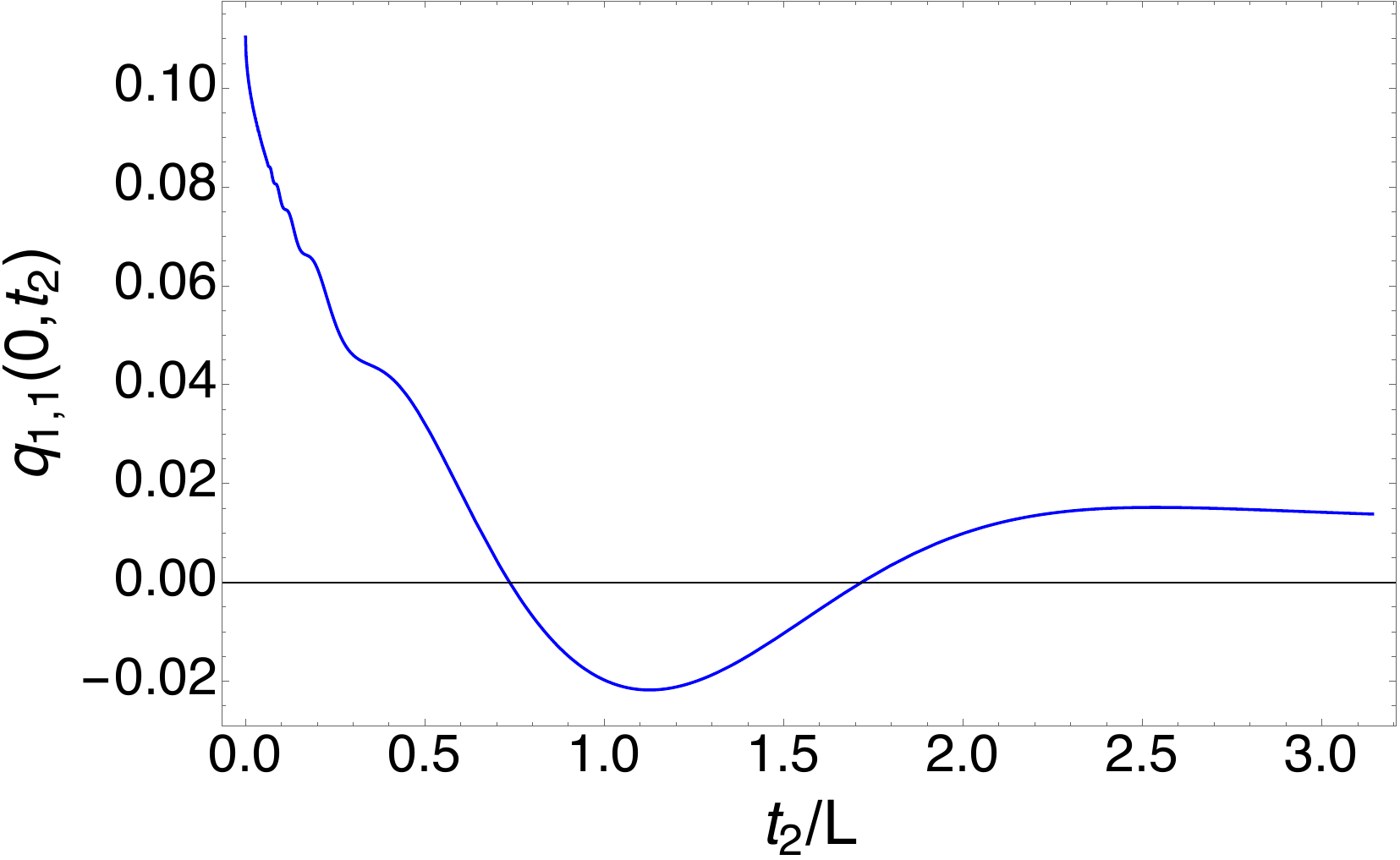}
    \hspace{0.5cm}
    \includegraphics[width=8cm]{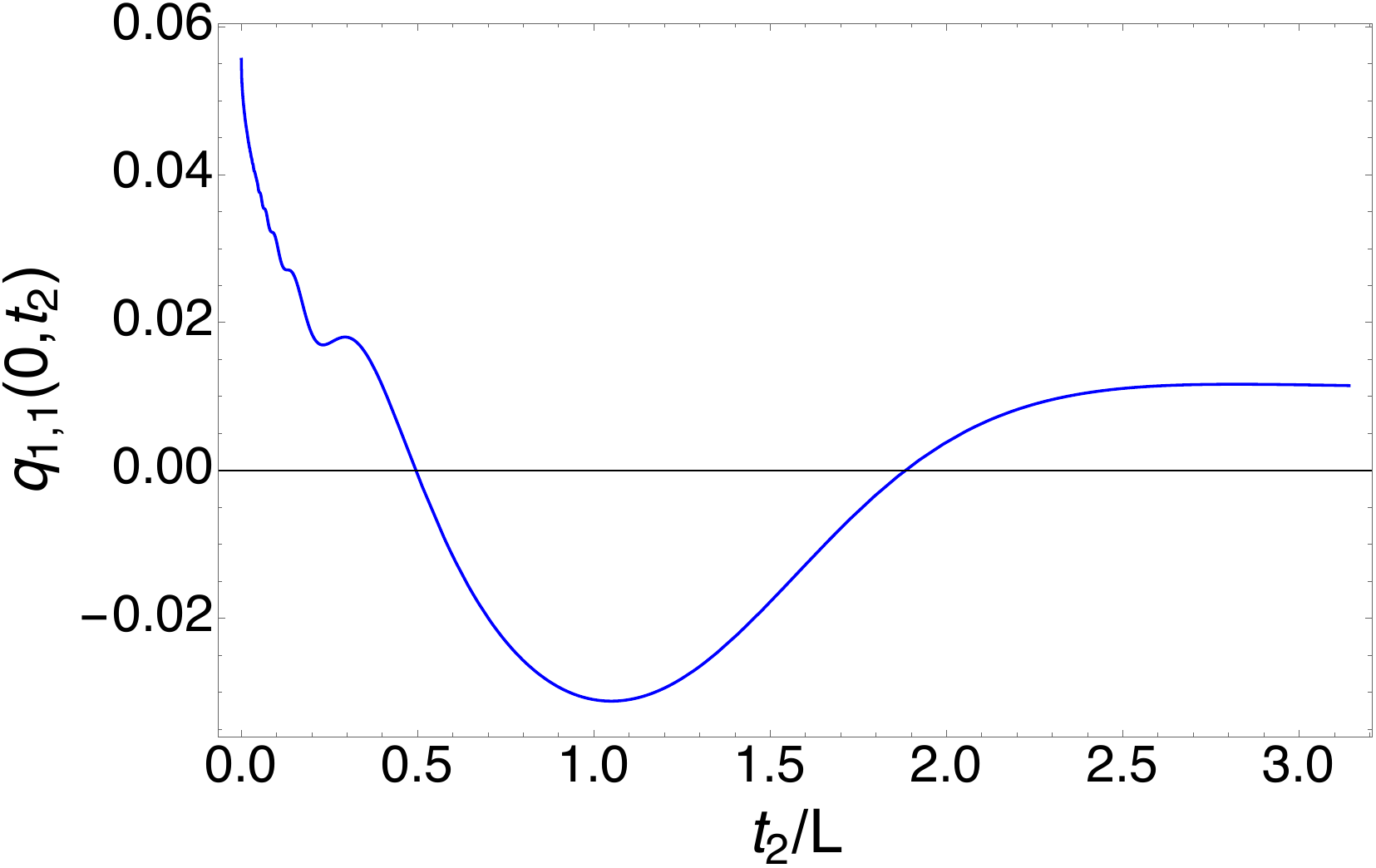}
    \caption{$q_{1,1}(0,t_2)$ as a function of $t_2/ L$ for the one-dimensional chiral massless field in the vacuum state and squeezed state with the projection  operator Eq.~(\ref{PSoned}). Here we fixed $r=0$ and $wL=0.45$ (left panel) and $r=0.3$ and $wL=0.4$ (right panel).
    }
    \label{fig:yamasakiondC}
\end{figure}
Figure~\ref{fig:yamasakiond}
plots the contour of the quasi-probability distribution function $q_{1,1}(0,t_2)$ on the plane of $w L$ and $ t_2/L$ for the vacuum state $r=0$ (left panel) and for the squeezed state with $r=0.3$ (right panel). One can see similar behaviors to the results of the three-dimensional model in Sec.~II C (see Figs.~\ref{fig:box}, \ref{fig:boxv}).
Similarly, figure~\ref{fig:yamasakiondB} plots $q_{1,1}(0,t_2)$
on the plane of $r$ and $t_2/L$. 
The quasi-probability distribution function has negative values smaller than $-0.02$ around $t_2/L\simeq 1.2$ and $wL\simeq 0.4\sim0.45$ even for the vacuum state.  
Figure \ref{fig:yamasakiondC} plots 
$q_{1,1}(0,t_2)$ as a function of $t_2/L$
for the model $wL=0.45$ and $r=0$ (left panel) and 
$wL=0.4$ and $r=0.3$ (left panel).
For the vacuum state and the squeezed state, the expectation values of the coarse-grained value
${\rm Tr}[\hat {\bar\phi}{}'(t)\rho_0]$ is always zero. On the other hand,
$q_{1,1}(0, t_2)$ is the quasi-probability that the measurement at $t_1 (= 0)$ gives $| {\bar\phi}{}'(0)| > w$ and the measurement at $t_2$
gives $| {\bar\phi}{}'(t_2)| > w$. This is the counter-intuitive result of measurements against the expectation values, which may occur
due to a spread of wave function and the superposition principle in quantum mechanical systems.

\vspace{0cm}
\section{Summary and conclusions}

We examined the violation of the Leggett-Garg inequalities for a quantum scalar field in a coherent state, the squeezed state with a coherent mode excitation as well as the squeezed state and the vacuum state, where we constructed the dichotomic variable of a coarse-grained field. We found that the violation of the Leggett-Garg inequalities may occur when $\xi\simgt 7$ and 
$1\simlt\ell L\simlt 3.5$ for the $(3+1)$ dimensional quantum field in one-mode coherent state. 
A similar violation appears for the $(1+1)$ dimensional chiral massless field, though the conditions for the violation are slightly different. 
We also demonstrated that the model of the one-mode coherent state with $\varphi(t)=0$ is equivalent to the model of the vacuum state choosing $\varphi(t)=-E(t)$. 
Further, we demonstrated that a simple choice of the dichotomic variable and the projection operator exhibits the violation of the Leggett-Garg inequalities for the vacuum state, as in Sec.~II D and Sec.~III C. 
Thus, by constructing the dichotomic variable properly, the violation of the Leggett-Garg inequalities can be observed for a quantum field in the vacuum state as well as the squeezed state. 
The violation of the Leggett-Garg inequalities occurs 
when the measurements give the opposite values to the expectation values, which can be understood in an analogy to a harmonic oscillator's model \cite{Hatakeyama}. 
The periodic behavior in the quasi-probability distribution function, which appears in the case of a harmonic oscillator \cite{Hatakeyama}, does not appear for a coarse-grained quantum field. This could be understood as a kind of decoherence effect coming from the fact that the dichotomic variable of the coarse-grained quantum field has a quantum correlation with the other region.
Thus our finding demonstrates a possible use of the Leggett-Garg inequalities for testing the quantum nature of a field.
We demonstrated the violations of the Leggett-Garg inequalities for the chiral massless quantum field in one dimension, 
which might be applied for an experiment of quantum Hall system \cite{Yusa,Hotta,Nambu}. 
In the present work, we didn't consider any physical process of measurements of the dichotomic variable, which is constructed by a coarse-grained quantum field in a finite local region. 
Predictions connected to such a realistic experiment are left in the future investigation (e.g., \cite{Hirotani}). 


\acknowledgements
We thank Yasusada Nambu, Akira Matsumura, Tomoya Hirotani, Youka Kaku, Yuki Osawa, Masahiro Hotta, Go Yusa, Satoshi Iso for the discussions related to the topic in the present work.
K.Y. was supported by JSPS KAKENHI (Grant No. JP22H05263 and No. JP23H01175).
D. M. was supported by JSPS KAKENHI (Grant No. JP22J21267).
 \vspace{0.1cm}
\appendix
\section{Useful formula for the quasi-probability distribution function}
Using the variables $c_1=c\cos{u},c_2=c\sin{u}$, 
the right-hand side of Eq.~(\ref{qssttsq}) leads to
\begin{align*}
    &q_{s_1,s_2}(t_1,t_2)
   \nonumber \\
    &~~~~= {\rm Re}\biggl[\frac{1}{4\pi(A_{\text{sq}}(t_1)A_{\text{sq}}(t_2)-B_{\text{sq}}(t_1,t_2)^2)^{{3}/{2}}}   \int_0^{{\pi}/{2}}du 
   \exp{-\frac{A_{\text{sq}}(t_2)E(t_1)^2+A_{\text{sq}}(t_1)E(t_2)^2-2B_{\text{sq}}(t_1,t_2)E(t_1)E(t_2)}{2(A_{\text{sq}}(t_1)A_{\text{sq}}(t_2)-B_{\text{sq}}(t_1,t_2))^2}}\nonumber\\
   &~~~~~~\biggr(\frac{2(A_{\text{sq}}(t_1)A_{\text{sq}}(t_2)-B_{\text{sq}}(t_1,t_2)^2)^2}{A_{\text{sq}}(t_2)\cos^2{u}+A_{\text{sq}}(t_1)\sin^2{u}-s_1s_2\sin{2u}}-\sqrt{2\pi}\biggr(\frac{A_{\text{sq}}(t_1)A_{\text{sq}}(t_2)-B_{\text{sq}}(t_1,t_2)^2}{A_{\text{sq}}(t_2)\cos^2{u}+A_{\text{sq}}(t_1)\sin^2{u}-s_1s_2B_{\text{sq}}(t_1,t_2)\sin{2u}}\biggr)^{{3}/{2}}\nonumber\\
   &~~~~~~\exp{\frac{(s_1A_{\text{sq}}(t_2)E(t_1)\cos{u}+s_2A_{\text{sq}}(t_1)E(t_2)\sin{u}-B_{\text{sq}}(t_1,t_2)(s_2\sin{u}E(t_1)+s_1\cos{u}E(t_2)))^2}{2(A_{\text{sq}}(t_1)A_{\text{sq}}(t_2)-B_{\text{sq}}(t_1,t_2)^2)(A_{\text{sq}}(t_2)\cos^2{u}+A_{\text{sq}}(t_1)\sin^2{u}-s_1s_2\sin{2u}B_{\text{sq}}(t_1,t_2)))}}\nonumber \\
   &~~~~~~{\rm erfc}\biggr[\frac{s_1A_{\text{sq}}(t_2)E(t_1)\cos{u}+s_2A_{\text{sq}}(t_1)E(t_2)\sin{u}-B_{\text{sq}}(t_1,t_2)(s_2\sin{u}E(t_1)+s_1\cos{u}E(t_2))}{2(A_{\text{sq}}(t_1)A_{\text{sq}}(t_2)-B_{\text{sq}}(t_1,t_2)^2)}\nonumber\\ 
   &~~~~~~\sqrt{\frac{2(A_{\text{sq}}(t_1)A_{\text{sq}}(t_2)-B_{\text{sq}}(t_1,t_2)^2)}{A_{\text{sq}}(t_2)\cos^2{u}+A_{\text{sq}}(t_2)\sin^2{u}-2s_1s_2B_{\text{sq}}(t_1,t_2)\sin{2u})}}\biggr] \nonumber \\
   &~~~~~~ \bigr(s_1A_{\text{sq}}(t_2)E(t_1)\cos{u}+s_2A_{\text{sq}}(t_1)E(t_2)\sin{u}-B_{\text{sq}}(t_1,t_2)(s_2E(t_1)\sin{u}+s_1E(t_2)\cos{u})\bigr)\biggr)\biggr],
\end{align*}
where ${\rm erfc}(z)$ is the complementary error function. 

\vspace{0.5cm}

\end{document}